\documentclass[12pt]{article}

\usepackage{graphics,psfrag,rotating,a4wide}
\usepackage{cite} 
\usepackage{epsfig,amssymb}
\usepackage{amsmath}
\newenvironment{Eqnarray}%
     {\arraycolsep 0.14em\begin{eqnarray}}{\end{eqnarray}}
                                                                                
\makeatletter
\@addtoreset{equation}{section}

\makeatother
                                                                                
\begin{document}
\def\nicefrac#1#2{\hbox{${#1\over #2}$}}

\newcommand{\nn}{\nonumber}
\newcommand{\raw}{\rightarrow}
\newcommand{\be}{\begin{equation}}
\newcommand{\ee}{\end{equation}}
\newcommand{\bea}{\begin{Eqnarray}}
\newcommand{\eea}{\end{Eqnarray}}
\newcommand{\dl}{\stackrel{\leftarrow}{D}}
\newcommand{\dr}{\stackrel{\rightarrow}{D}}
\newcommand{\dd}{\displaystyle}
\newcommand{\slas}[1]{\rlap/ #1}
\newcommand{\rmd}{{\rm d}}

\def\lsim{\mathrel{\raise.3ex\hbox{$<$\kern-.75em\lower1ex\hbox{$\sim$}}}}
\def\gsim{\mathrel{\raise.3ex\hbox{$>$\kern-.75em\lower1ex\hbox{$\sim$}}}}
\def\ifmath#1{\relax\ifmmode #1\else $#1$\fi}
\def\beq{\begin{equation}}
\def\eeq{\end{equation}}
\def\beaa{\begin{array}}
\def\eeaa{\end{array}}
\def\ts{\tilde{t}}
\def\cs{\tilde{c}}
\def\sabd{$ \sin \left(\beta - \alpha \right)$ }
\def\cabd{$ \cos \left(\beta - \alpha \right)$ }
\def\sab{ \sin \left(\beta - \alpha \right) }
\def\cab{ \cos \left(\beta - \alpha \right) }
\def\sba{ \sin \left(\beta - \alpha \right) }
\def\cba{ \cos \left(\beta - \alpha \right) }
\def\ls#1{\ifmath{_{\lower1.5pt\hbox{$\scriptstyle #1$}}}}
  
\pagestyle{empty}

\begin{flushright}
MPP-2004-33\\
CERN-PH-TH/2004-072\\
hep-ph/0511021\\
\end{flushright}

\renewcommand{\thefootnote}{\fnsymbol{footnote}}
\begin{center}
{\Large\bf
Flavour changing effects on 
$e^{+} e^{-}\rightarrow H b \bar s, \, H \bar b s$  \\[0.3cm]
in the MSSM}\\[1.3cm]
{\large 
W.~Hollik$^{1}$, S.~Pe{\~n}aranda$^{2}$ and M.~Vogt$^{1}$
~\footnote{electronic addresses:\\
\hspace*{0.6cm} hollik@mppmu.mpg.de,\\ 
\hspace*{0.6cm} siannah.penaranda@cern.ch,\\ 
\hspace*{0.6cm} mvogt@mppmu.mpg.de}
}\\[7pt]
$^1$ {\it Max-Planck-Institut f\"ur Physik, F\"ohringer Ring 6, 
D--80805 Munich, Germany}\\[1.5pt]
$^2$ {\it CERN-TH Division, Department of Physics, CH-1211 Geneva 23, Switzerland}
\\[1cm]

\begin{center}
{\bf Abstract}
\end{center}
\end{center}

Flavour changing effects originating
from the exchange of scalar particles in
 the processes $e^{+} e^{-}\rightarrow H^x b \bar s,\, H^x \bar b s$, with
 $H^x\equiv h^0\,, H^0\,, A^0$, are investigated
in the Minimal Supersymmetric Standard
 Model with non-minimal flavour violation at the one-loop level.
The dominating SUSY-QCD contributions with
squark--gluino loops are calculated and discussed.
We consider the SUSY scenario with non-minimal flavour mixing
in the down-type squark-mass matrix.
The flavour-changing cross sections are derived, and we discuss the 
dependence on the MSSM parameters and the strength of flavour mixing.
The values for the cross section can reach $10^{-4}$ pb  for the 
production of the heavy Higgs boson $H^0$ or $A^0$, and  
only $10^{-7}$ pb for the light Higgs boson $h^0$. 
Non-decoupling behaviour occurs for both $h^0, H^0$ production
in the case of a common heavy SUSY mass scale.
     
\vfill
\clearpage

\renewcommand{\thefootnote}{\arabic{footnote}}
\setcounter{footnote}{0}

\pagestyle{plain}
\section{Introduction}
\label{chap.introduction}
 
Searching for Flavour Changing Neutral Currents (FCNC) is among the 
important tasks of the coming generation of high energy
colliders~\cite{report}. FCNC processes are forbidden 
in the Standard Model (SM) at lowest order.
These effects appear at the loop level and are of basic importance
for testing the quantum structure of the SM. 
In the SM, however, the one-loop effects are small, 
suppressed by the GIM mechanism~\cite{GIM}.
In models beyond the SM new non-standard particles 
appear in the loops, with significant contributions to
flavour changing transitions~\cite{FCMSSM}. 
Therefore, FCNC processes play an
important role in searching new physics beyond the SM.

Among various new physics models, supersymmetry (SUSY),
especially the Minimal Supersymmetric Standard Model (MSSM), is a
favoured candidate~\cite{nilles}. The MSSM introduces two
Higgs doublets to break the electroweak symmetry. After symmetry
breaking, there are five physical Higgs bosons: two CP-even Higgs
bosons ($h^0\,, H^0$), one CP-odd boson ($A^0$)
and two charged Higgs bosons ($H^\pm$)~\cite{HHG}. 
The couplings of these Higgs particles may differ significantly from
those of the SM. In fact, an important feature of SUSY models is that
the fermion-Higgs couplings are no longer strictly proportional to the
corresponding mass, as they are in the SM. In particular, the b-quark 
coupling to the neutral Higgs boson becomes enhanced for large 
$\tan \beta=v_2/v_1$, the ratio of the two
vacuum expectation values~\cite{HHG}~{\footnote{Explicitly they are proportional to $1/\cos \beta$, and to $[-\sin \alpha\,, \cos \alpha, \sin \beta]$ for
$[h^0\,, H^0\,, A^0]$ respectively.}}. 
Since there are five Higgs bosons in
the MSSM, additional features not present in the SM may be useful to pin
down differences between those models, manifested in different coupling
strengths, decay widths and production cross sections.
In the minimal flavour violation scenario (MFV), the only source of
flavour violation is mediated by the CKM-Matrix~\cite{FCMSSM}. 
In the general MSSM 
with non-minimal flavour violation (NMFV), new flavour changing (FC)
effects emerge due to a possible misalignment between the squark and 
the quark sector. The FC interactions resulting from such a 
misalignment do not show up at the tree-level, but
they can be generated at the one-loop level and could lead to relevant 
contributions to observables for specific regions of the MSSM parameters.
The MSSM with NMFV is considered in this paper.

The FCNC vertices in the SM and beyond have been extensively examined in
the literature, and the results promise FCNC to provide a fertile ground for 
testing the SM and probing new physics. These analysis include rare 
decays of $B$-meson systems~\cite{hadrons,buras}, Z-boson
decays~\cite{ZSM,Zbs,otrosZ}, and top decays~\cite{top}.
It is known that the SM predictions for the top quark FCNC processes are
far below the detectable level and that the MSSM can enhance them by several
orders, making them potentially accessible at future collider
experiments~\cite{joseantonio}. In particular,
the branching ratios for $Z \rightarrow b \bar s$
decays are of the order of $10^{-8}$ in 
the SM~\cite{ZSM}, of the
same order in SUSY with $\tilde t$-$\tilde c$ mixing and can 
reach $10^{-6}$ in SUSY with $\tilde b$-$\tilde s$ mixing~\cite{Zbs},
both last rates being dominated by the SUSY radiative effects from
squark-gluino loops. FC effects that can be induced by
squark-gluino loops in Higgs-boson decays have
been investigated as well, and 
$BR(H^x \rightarrow b \bar s) \sim 10^{-4}-10^{-3}$
and $BR(H^x  \rightarrow t \bar c) \sim 10^{-4}$ have been 
found for selected regions of the MSSM
parameter space and the flavour mixing parameters in the 
MSSM squark-mass matrices~\cite{maria,sola}. The electroweak
corrections are subdominant, at least one order of magnitude 
smaller than the SUSY-QCD ones, but can give interference effects~\cite{florian}. 
The large rates found for the SUSY contributions to the Higgs partial 
decay widths into $b \bar s$ and $s \bar b$, as well as to the effective 
FC Higgs couplings to quarks, are indeed quite 
encouraging~\cite{maria,sola,florian,kolda,isidori,dedes,Demir}.
Furthermore, effects of non-minimal flavour violation on the MSSM Higgs
boson masses and the electroweak precision observables, $m_W$ and $\sin
\theta_{eff}$, has been investigated in~\cite{sven}. 

FC effects in the SM in electron--positron collisions have been 
under study in several papers.
In particular, the cross sections 
of $e^+e^-\rightarrow t\bar c$ and $e^+e^-\rightarrow b\bar s$ processes
in the SM are analyzed in~\cite{chino}.  The cross section of the
process with a $b\bar s$ final state was found to be larger
than the one with a $t \bar c$ final state, but only of the order 
of $10^{-3}$ fb, which is too small to be observed at present and future 
colliders. 

In this paper we investigate FC effects
via  associated bottom-strange and Higgs boson production in the
MSSM. More concretely, we study the 
effects of squark-gluino loops in the higher-order processes 
\begin{eqnarray}
e^+e^-\rightarrow H^x b\bar{s} + H^x \bar{b}s\,\hspace*{1.0cm}
(H^x\equiv h^0\,,H^0\,, A^0) \,,
\end{eqnarray}
and discuss the size of the induced cross sections
at the one-loop level, with virtual
second and third generation squarks,
and their dependence on the 
MSSM parameters and the flavour mixing strength.

The paper is organized as follows: 
In section~\ref{chap.introd.} we present a brief outline on 
squark mixing in the MSSM with NMFV.
Section~\ref{chap.analitical} 
gives an overview over the various classes of 
squark--gluino diagrams at the one-loop level and their impact
on the total cross section,
including both resonant diagrams (subsection~\ref{chap.resonances})
and non-resonant contributions (subsection~\ref{chap.phenomenology}). 
We discuss the dependence on the MSSM parameters in 
section~\ref{chap.numerical}. 
Additionally, the non-decoupling behaviour of the SUSY contributions
in the large sparticle-mass limit is presented in 
section~\ref{chap.decoupling}, and a summary is given in 
section~\ref{conclu}.

\section{Non-minimal squark mixing}
\label{chap.introd.}

Flavour changing phenomena in SUSY models can emanate from a mixing 
between different generations of squarks through the soft breaking 
terms in the Lagrangian of the squark sector. 
This non-minimal flavour mixing scenario, where squark mixing results
from a misalignment between the quark and the squark mass matrices, is the most
general case of the MSSM. We assume that the flavour changing squark
mixing is significant only in transitions between third and second
generation squarks. It is known that there are strong experimental
bounds on the squark mixing involving the first generation, 
resulting from data on $K^0-\bar K^0$ and 
$D^0-\bar D^0$~mixing~\cite{FCMSSM}.

In general, the flavour violating quantities arise from
non-diagonal entries in the bilinear soft breaking matrices 
$M_{\tilde Q}^2\,, M_{\tilde U}^2$ and $M_{\tilde D}^2$, that appear in
the mass matrices in the up-squark and down-squark sectors as well as
from non-diagonal (flavour changing) entries in the trilinear soft
breaking matrices $A_u$ and $A_d$. 
In our analysis, FC effects are generated by the one-loop
exchange of $\tilde{b}$-$\tilde{s}$-admixture states and gluinos,
$\tilde g$. When the squark mass matrix is diagonalized, FC
gluino-quark-squark couplings can be derived for the mass eigenstates and
FC effects via squark-gluino loops are induced. 
It is known that some of the flavour mixing parameters in the
$\tilde{b}-\tilde{s}$ sector are severely constrained by the
$b\rightarrow s \gamma$ decays~\cite{FCMSSM,isidori,bsgamma,bsgamma2}, 
but in particular the ones referring to the LL- and RR-mixing 
of the SUSY partners of the left-handed and right-handed quarks,
respectively, are not definitely excluded~\cite{FCMSSM}.

We assume the simplest case of mixing, 
where the only non-zero off-diagonal squark 
squared mass entries in the squark-sector stand for 
$\tilde s_L \tilde b_L$ and $\tilde c_L \tilde t_L$  mixing. 
Expressing the squark mass matrix in terms of the chirality eigenstates 
$\{ s_L, s_R, b_L, b_R  \}$ and $\{ c_L, c_R, t_L, t_R  \}$, 
respectively, and considering only LL-Mixing, i.e.\ mixing between the 
left-chiral components of the squarks leads to the following mass matrix 
mediating between the chiral eigenstates and the down-squark 
admixtures~\cite{maria,florian},
\begin{eqnarray}
M^2_{\tilde u} =\left\lgroup
         \beaa{llll}
          M_{L,c}^2  &   m_c X_c & \lambda_{LL} M_{L,c} M_{L,t}& 0\\
           m_c X_c   &  M_{R,c}^2  & 0  &  0\\
          \lambda_{LL} M_{L,c} M_{L,t} & 0 & M_{L,t}^2  &   m_t X_t\\
          0 & 0 &  m_t X_t &  M_{R,t}^2
\eeaa
         \right\rgroup ,
\label{eq.usquarkmass}
\end{eqnarray}
\begin{eqnarray}
M^2_{\tilde d} =\left\lgroup
         \beaa{llll}
          M_{L,s}^2  &   m_s X_s & \lambda_{LL} M_{L,s} M_{L,b}& 0\\
           m_s X_s   &  M_{R,s}^2  & 0  &  0\\
          \lambda_{LL} M_{L,s} M_{L,b} & 0 & M_{L,b}^2  &   m_b X_b\\
          0 & 0 &  m_b X_b &  M_{R,b}^2
\eeaa
         \right\rgroup
\label{eq.dsquarkmass}
\end{eqnarray}
with
\begin{eqnarray}
M_{L,q}^2 &=& M_{\tilde Q,q}^2 +m_q^2 + \cos2\beta (T_3^{q}-Q_q s_W^2)m_Z^2\, , \nn \\
M_{R,(c,t)}^2 &=& M_{\tilde U,(c,t)}^2 +m_{c,t}^2 + \cos2\beta Q_t s_W^2 m_Z^2\, , \nn \\
M_{R,(s,b)}^2 &=& M_{\tilde D,(s,b)}^2 +m_{s,b}^2 + \cos2\beta Q_b s_W^2 m_Z^2\, , \nn \\
X_{c,t} &=& A_{c,t} - \mu \cot \beta \, , \nn \\
X_{s,b} &=& A_{s,b} - \mu \tan \beta \, ,
\label{eq.squarkparam}
\end{eqnarray}
where $M_{L,q}$, $M_{R,q}$ are the corresponding bilinear soft SUSY
breaking entries and $m_q$ the quark matrices with $q \in \{b,s, c, t
\}$. $T_3^q$ is the third component of the weak isospin, $Q_q$ denotes
the charge of the corresponding quark. The trilinear soft SUSY breaking
matrices $A_q$ are assumed to be diagonal, with entries $A_{s,b}$. $\mu$
is the mass parameter of the Higgs boson sector and $\tan \beta$ is the
ratio of the vacuum expectation values in this sector. $m_Z$ is 
the $Z$ boson mass, and $s_W = \sin\theta_W$ contains
the electroweak mixing angle $\theta_W$. $\lambda \equiv \lambda_{LL}$ denotes the 
parameter characterizing the flavour mixing strength in the mixing of 
the LL-components of the second and third generation
squarks~\footnote{The flavour mixing  parameter $\lambda$ can be 
identified with 
  $(\delta_{LL})_{23}$ in the usual notation of the mass-insertion 
approximation~\cite{FCMSSM}.}. Thus, 
we have only one free parameter determining the strength of FC.

\section{The loop-induced cross sections}  
\label{chap.analitical}

In this section we discuss the various contributions leading to the
production of a Higgs boson associated with a flavor-nondiagonal
quark pair, $b \bar s$ or $s \bar b$.
As far as the final states with  $b \bar s$ and $s \bar b$ are experimentally
not distinguished, the result for 
$\sigma (e^{+} e^{-} \rightarrow H^x \, b\, \bar s+ H^x \,s\, \bar b)$
is obtained by summing the two individual cross sections. 

We have done the computation using {\it FeynArts}, {\it FeynCalc} and 
{\it FormCalc}~\cite{hahn}. Feynman rules
of MSSM vertices with FC effects are recently implemented in {\it FeynArts}
(extending the previous MSSM model file).
The Higgs boson masses, the  Higgs boson decays and
the masses of the MSSM-spectrum have been computed using 
the {\it FeynHiggs2.1beta} Fortran code~\cite{feynhiggs}, including all
FC effects.
As the MSSM Higgs boson couplings to down-type fermions receive
large quantum corrections enhanced by $\tan \beta$, these corrections
have to be resummed to all orders in perturbation theory with the help of
the effective Lagrangian formalism~\cite{Carena}. 
These leading threshold corrections to the bottom mass
are taken into account in our computation in terms of the resummed version.

\subsection{Kinematics}

The cross section for the production process
\bea
e^-(p_a) + e^+(p_b) \rightarrow\, b(p_1) + \bar s(p_2) + H^x(p_3)
\eea
is given by the expression (neglecting the electron mass)
\bea
\sigma  = \frac{1}{2 s (2 \pi)^5}  
\int \prod_{i=1}^{3}\left[ \frac{\rmd^3 p_{i}}{2p_i^0}\right] \delta^4
(p_a + p_b - \sum_{i} p_{i})\,  \overline{|{\cal M}|^2} \, ,
\eea
involving an integration over the 5-dimensional phase space and
$\overline{|{\cal M}|^2}$,
the spin-averaged matrix-element squared of the process.
The phase space integral
can be expressed in terms of two energy  and two angular variables, 
chosen in the overall CMS with $s=(p_a+p_b)^2$,
\bea
\sigma  =  \frac{1}{16 s (2 \pi)^{4}} 
\int^{p^+_1}_{m_1} \rmd p^0_1  
\int^{p^+_2}_{p^-_2} \rmd p^0_2
\int^{+1}_{-1} \rmd\cos\theta \int^{2 \pi}_0 \rmd\eta \,\, 
       \overline{|{\cal M}|^2} \, , 
\eea
with the boundaries
\bea
 p_1^+  &=& \frac{\sqrt{s}}{2} - \frac{(m_2 + m_3)^2 - m_1^2}{2 \sqrt{s}}\, , \nn \\
 p_2^{\pm} &=& \frac{1}{2 \tau} \left[ \sigma (\tau +  m_{+} m_{-}) 
\pm \lvert \vec{p}_1 \rvert \sqrt{(\tau - m_{+}^2)(\tau - m_{-}^2)} \right] ,
\eea
where
\bea
\sigma = \sqrt{s} - p^0_1 , \, \, \tau = 
\sigma^2 - \lvert \vec{p}_1 \rvert^2, \, \,  m_{\pm} = m_2 \pm m_3 \, . \nn
\eea
$\theta$ is the angle between the momenta $\vec{p}_a$ and $\vec{p}_1$,
and $\eta$ the angle between the two planes spanned by 
$\vec{p}_1,\vec{p}_2$ and  $\vec{p}_1,\vec{p}_a$.

\subsection{Resonant contributions}
\label{chap.resonances}

Due to the structure of the couplings
of Higgs bosons to another Higgs boson and/or to a Z-boson in the MSSM, 
the $q \bar q H$
production processes receive large resonating contributions not
present in the SM. The associated Higgs boson production with a heavy 
quark pair, $t\bar t$ and $b\bar b$, in $e^{+}  e^{-}$ collisions at 
high energy have been investigated and large resonant contributions that
imply rates of a few fb  have been found~\cite{dawson} for the 
$b\bar b$-pair. Here we consider the resonating contributions to the
associated Higgs boson production with a quark pair $b\bar s$ in $e^{+}
e^{-}$ collisions. 

\begin{figure}[t!]\begin{center}
\rotatebox{270}{\epsfig{file=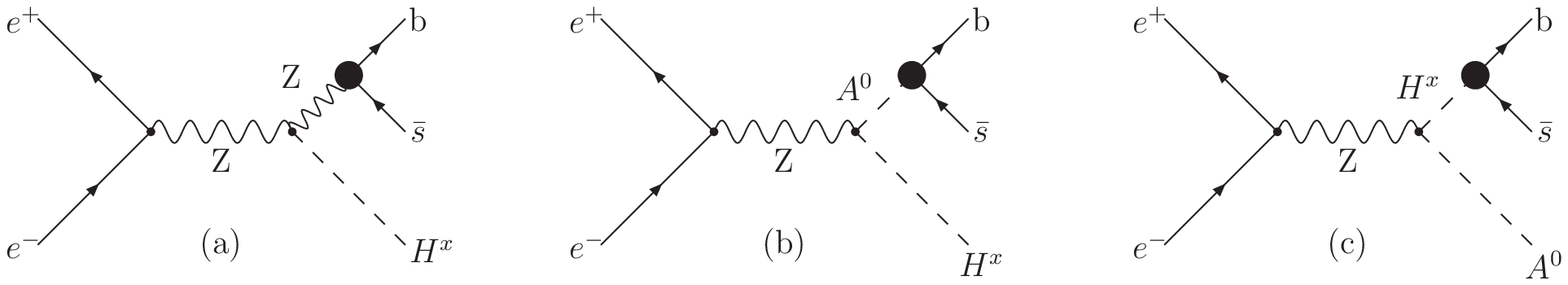,scale=0.72}}
\hspace*{-0.8cm}\epsfig{file=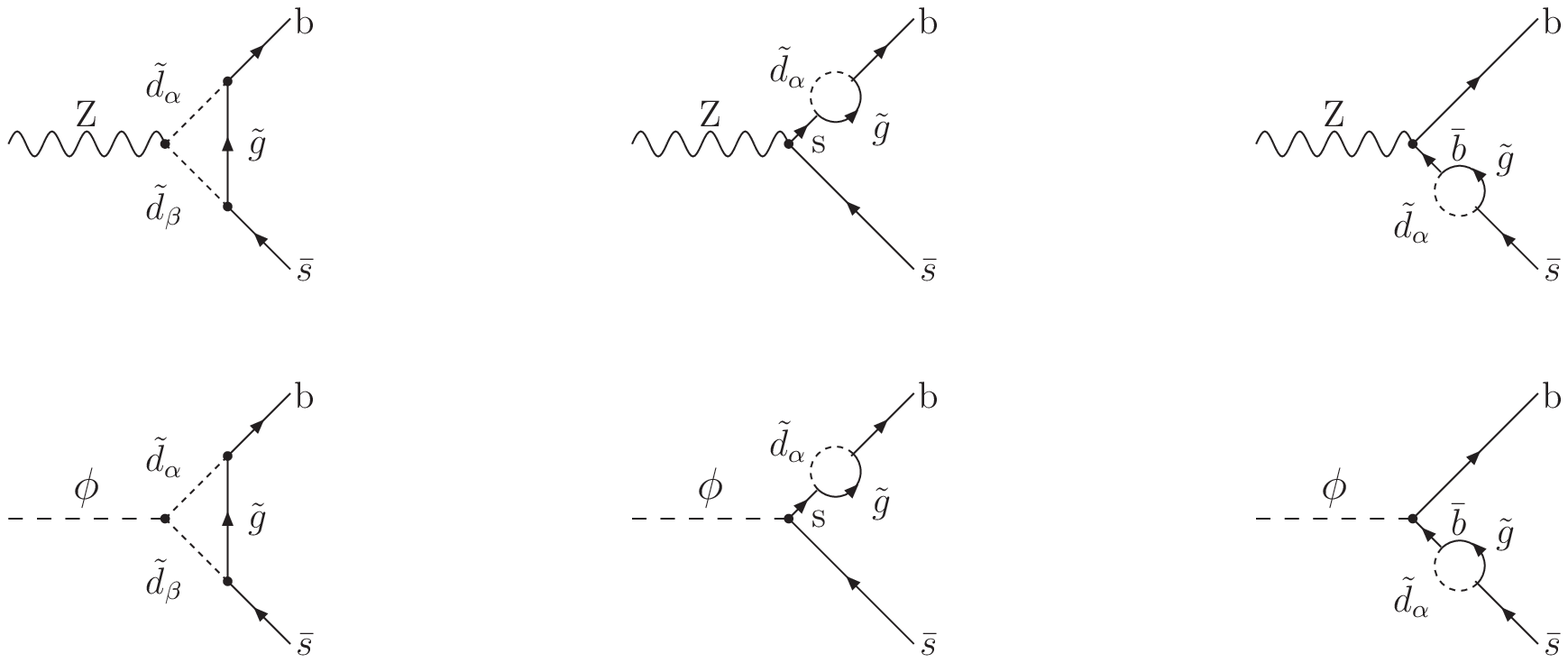,scale=0.72}\\
\vspace*{-0.8cm}{\footnotesize{(d)}}
\caption{{(a), (b), (c)} Generic diagrams for resonant contributions to
$\sigma (e^{+} e^{-} \rightarrow 
 \phi\, b\, \bar s+\phi \, s \,\bar b)$ $(\phi \equiv h^0\,, H^0\,, A^0)$.
{(d)} Gluino-squark loop contributions to $Z \rightarrow b\,
\bar s$ and $\phi \rightarrow b\, \bar s$.}
\label{diagsresonancia}\end{center}
\end{figure}

The structure of the corresponding Feynman diagrams is shown in
Fig.~\ref{diagsresonancia}; other diagrams are suppressed by
the small electron mass. In the kinematically allowed regions
there are two types of resonating intermediate states for $H^x=h^0,H^0$ 
production, involving $Z$ and $A^{0}$ resonances,
$e^{+} e^{-} \rightarrow Z H^x$ with $Z \rightarrow b \bar s$, and 
$e^{+}  e^{-} \rightarrow A^0 H^x$ with $A^0 \rightarrow b \bar s$
(Fig.~\ref{diagsresonancia}{(a,b)}).
There are other resonances channels for $A^0$ production, involving 
 $h^0$ and $H^0$ resonances; namely 
$e^{+}  e^{-} \rightarrow A^0 H^x$ with $H^x \rightarrow b \bar s$ 
(Fig.~\ref{diagsresonancia}{(c)}). 
The result for the $H^x=h^0$ case ($h^0$ resonance) is several
 orders of magnitude smaller than the $H^0$ resonance case (see below), 
therefore we do not include this case in the discussion. 
These processes are sensitive to the MSSM couplings of the 
$Z$ boson to the Higgs boson
$H^x$ and to the $A^0 H^x$ pair, normalized to the 
standard $ZZ h_{SM}$ coupling given by
\begin{eqnarray}
\label{eq:coup_res}
g_{Z A^0 h^0} = \cos (\beta -\alpha)\,, &\, g_{Z Z h^0}= \sin (\beta
-\alpha)\,,\nonumber\\
g_{Z A^0 H^0} = \sin (\beta -\alpha)\,, &\, g_{Z Z H^0}= \cos (\beta
-\alpha)\,.
\end{eqnarray}

For the computation, we treat the intermediate particle 
in the narrow-width approximation, i.e.\ multiplying the
two-particle production cross section for 
on-shell particles with the corresponding branching ratios,
\begin{eqnarray}
\sigma (e^{+} e^{-} \rightarrow 
H^x \, b\, \bar s+H^x \, s \,\bar b)&\simeq & \sigma (e^{+} e^{-} \rightarrow 
H^x \,Z) \cdot BR(Z \rightarrow  b\, \bar s +s \,\bar b)\,,\nonumber\\
\sigma (e^{+} e^{-} \rightarrow 
H^x \, b\, \bar s+H^x \, s \,\bar b)&\simeq & \sigma (e^{+} e^{-} \rightarrow 
H^x \,A^0) \cdot  BR(A^0 \rightarrow  b\, \bar s +s \,\bar b)\,,\nonumber\\
\sigma (e^{+} e^{-} \rightarrow 
H^x \, b\, \bar s+H^x \, s \,\bar b)&\simeq & \sigma (e^{+} e^{-} \rightarrow 
H^x \,A^0) \cdot  BR(H^x \rightarrow  b\, \bar s +s \,\bar b)\,,
\end{eqnarray}
which yields a good approximation owing to
the small total decay widths of the intermediate resonances.
The flavor-changing $Z$, $A^0$ and $H^x$ decays are loop-induced, 
with the gluino-squark strong contributions shown in 
Fig.~\ref{diagsresonancia}{(d)} as the dominating source,
yielding the branching ratios of $Y = Z, A^0, H^x$,
\bea
BR\, (\,Y \rightarrow  b\, \bar s+s \,\bar b\,) =
\frac{\Gamma \,(\,Y \rightarrow b\, \bar s\,) + \Gamma \,(\, Y \rightarrow s
  \,\bar b\,)}{\Gamma_{\rm tot}(Y)} \,,
\eea
where $\Gamma_{\rm tot}(Y)$ is the total width in each case.

Although numerical results and a discussion of the parameter dependence
will be given in the next section, we want to illustrate here
the size of the various contributions, choosing $m_A = 250$ GeV,
$\tan\beta =35$, and the SUSY parameters of (\ref{eq:numparameters}).
For the value $\lambda = 0.6$ of the mixing parameter, one obtains
the following branching ratios,
\begin{eqnarray}
&& BR(A^0 \rightarrow  b\, \bar s +s \,\bar b) \sim 10^{-2}\,,~~~~~~
 BR(H^0 \rightarrow  b\, \bar s +s \,\bar b) \sim 10^{-2}\,,\nonumber\\
&& BR(h^0 \rightarrow  b\, \bar s +s \,\bar b) \sim 10^{-4}\,,~~~~~~
BR(Z \rightarrow  b\, \bar s +s \,\bar b)\sim 10^{-11} \,.
\end{eqnarray} 
Here the threshold corrections on the $b$-quark mass are considered.
This implies that the results for $Z \rightarrow  b\, \bar s$ 
are different from the ones given in~\cite{Zbs}. 
For $m_b=5$ GeV we can reproduce the previous results.

The couplings given in~(\ref{eq:coup_res}) can
be suppressed if either $\sin (\beta-\alpha)$ or $\cos
(\beta-\alpha)$ is very small. 
In particular for large $\tan\beta$ and not too low $m_A$, as it is
the case for the choice of input parameters~(\ref{eq:numparameters}),
the $Z A^0 h^0$ and $Z Z H^0$ couplings are suppressed
and the couplings for $Z A^0 H^0$ and $Z Z h^0$ are large. Hence,
the cross sections for $e^{+} e^{-} \rightarrow 
H^0 \,A^0$ and $e^{+} e^{-} \rightarrow  h^0 \,Z$ are
about 2--3 orders of magnitude larger than for $e^{+} e^{-} \rightarrow 
h^0 \,A^0$ and $e^{+} e^{-} \rightarrow  H^0 \,Z$, at $\sqrt{s}=500$ GeV
amounting to
\begin{eqnarray}
 \sigma (e^{+} e^{-} \rightarrow H^0 \,A^0) \simeq 10^{-2} \,{\rm pb} \,\,,
& &
\sigma (e^{+} e^{-} \rightarrow h^0 \,Z) \simeq 10^{-2}-10^{-1} \,{\rm pb} , \nn
\\
 \sigma (e^{+} e^{-} \rightarrow h^0 \,A^0) \simeq 10^{-5}-10^{-4} \,{\rm pb} \,, 
& & 
\sigma (e^{+} e^{-} \rightarrow H^0 \,Z) \simeq 10^{-5}-10^{-3} {\rm pb} \,.
\end{eqnarray} 
Therefore, we can conclude that the $Z$-resonance contribution 
for the $H^x b\bar{s}$ final state ($H^x \equiv h^0\,, H^0\,, A^0$)
are very small, at most of the order of $10^{-12}$ pb, 
and the $h^0$-resonance contributions are of the order of $10^{-8}$ pb, 
whereas the contributions with heavy Higgs bosons ($A^0\,, H^0$) 
intermediate states can reach $10^{-4}$ pb for the cross section.

\psfrag{sigAh0}{{{\rotatebox{90}{$\sigma$ [pb]}}}}
\psfrag{sigAh0}{{\rotatebox{90}{{$\sigma$ [pb]}}}}
\psfrag{sigAH0}{{{\rotatebox{90}{$\sigma$ [pb]}}}}
\psfrag{sqrts}{{$\sqrt{s}$}}
\psfrag{sqrtsGeV}{{$\sqrt{s}$\, [GeV]}}
\psfrag{@GeVD}{{$\,\hspace*{0.4cm}\,$[GeV]}}
\psfrag{@pbD}{{$\,\hspace*{1.2cm}[\,pb\,]$}}
\begin{figure}[htb!]
\begin{center}\hspace*{-0.6cm}
\epsfig{file=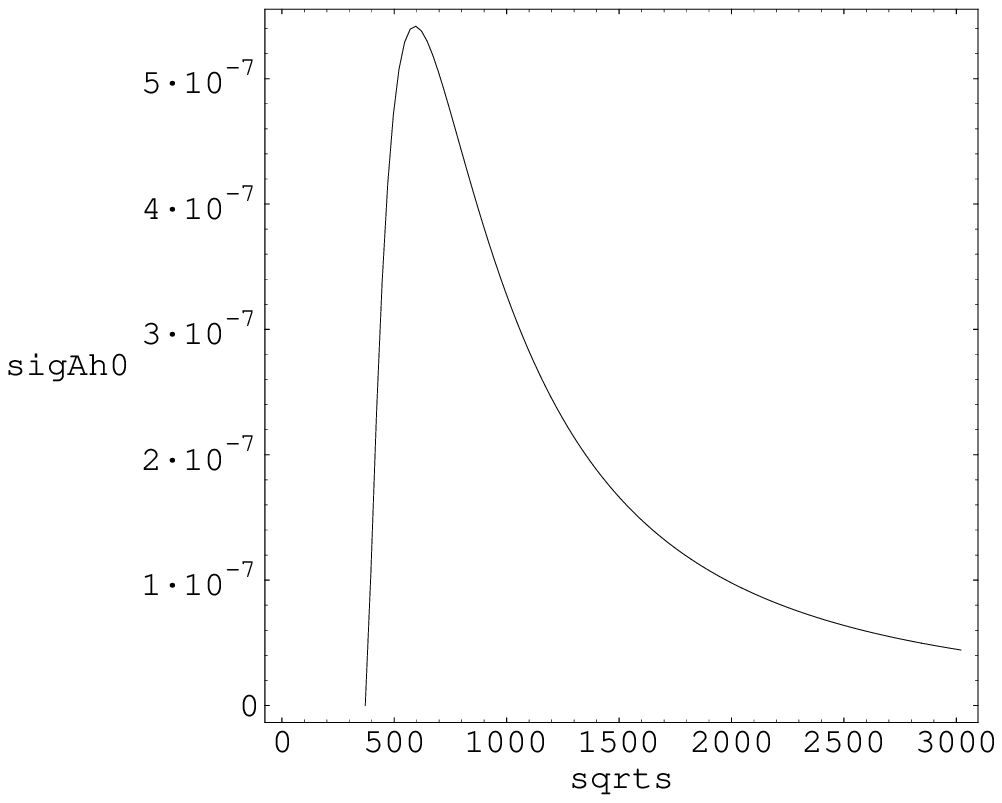,scale=0.74}
\epsfig{file=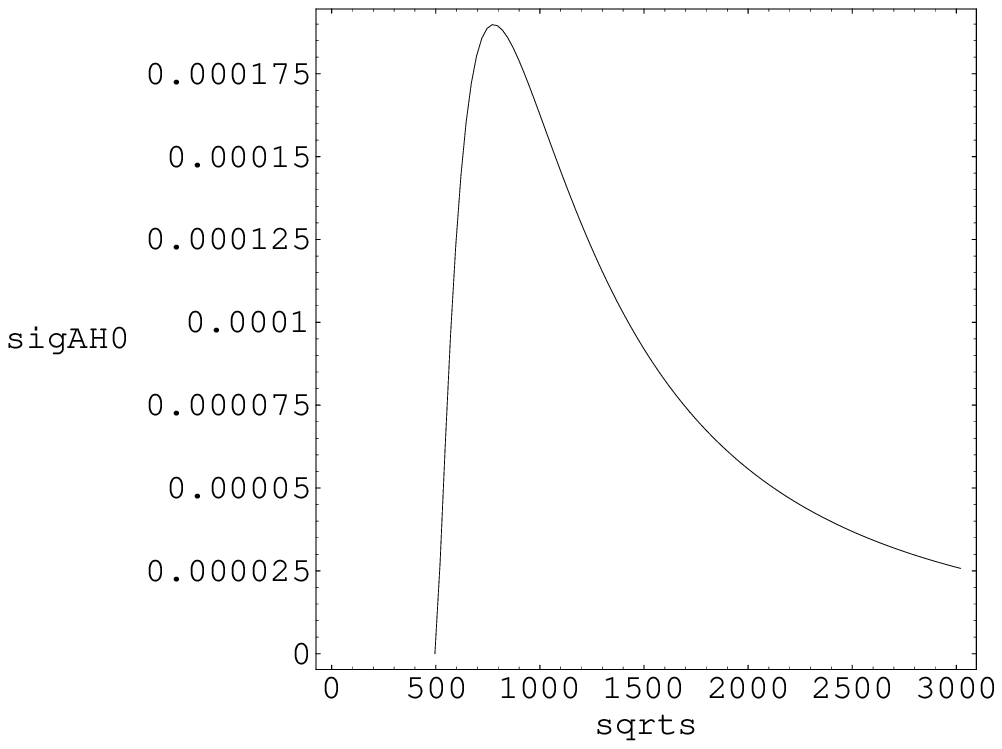,scale=0.80}
\caption{$A^0$-resonant contributions to $\sigma(e^{+} e^{-} \rightarrow 
H^x \, b\, \bar s+ H^x \,s\, \bar b)$ as a function of $\sqrt{s}$ [GeV] 
for $\,H^x \equiv h^0$ (left) and $H^x \equiv H^0$ (right), with $\lambda=0.6$.}
\label{resonance}
\end{center}\vspace*{-0.7cm}
\end{figure}
\psfrag{Lambda}{{$\lambda$}}
\psfrag{aa}{{\footnotesize{$\sqrt{s} = 500$ GeV}}}
\psfrag{bb}{{\footnotesize{$\sqrt{s} = 1000$ GeV}}}
\psfrag{cc}{{\footnotesize{$\sqrt{s} = 1500$ GeV}}}
\begin{figure}[htb!]\vspace*{-0.5cm}
\begin{center}
\raisebox{0.4cm}{\epsfig{file=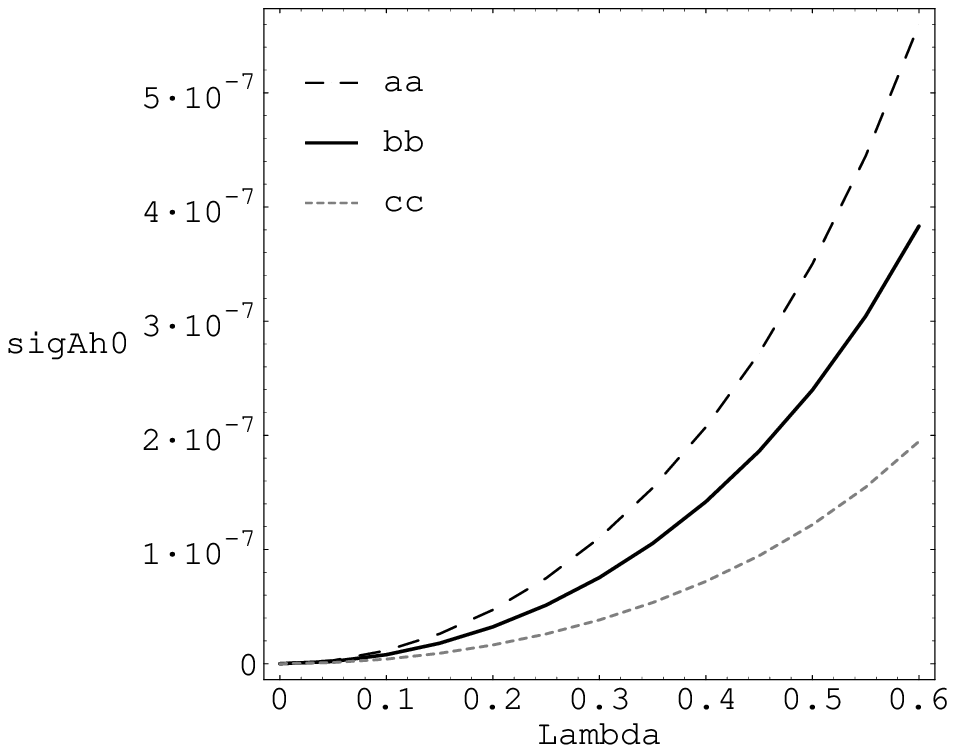,scale = 0.79}}\hspace*{-0.3cm}
\epsfig{file=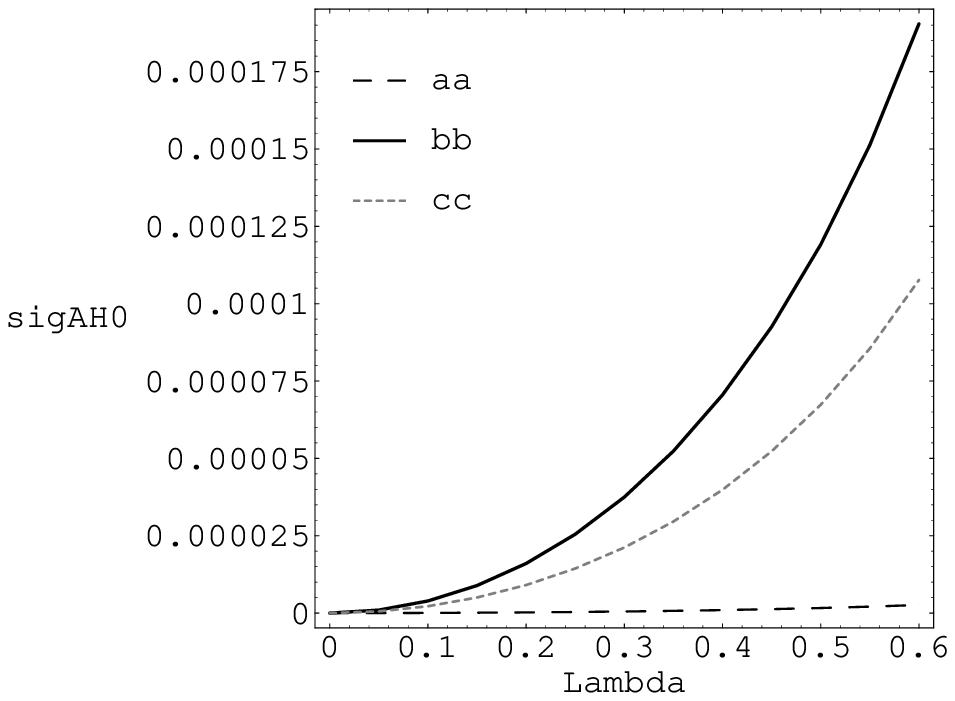,scale = 0.86}
\vspace*{-1.8cm}
\caption{Resonant contributions to $\sigma(e^{+} e^{-} \rightarrow 
H^x \, b\, \bar s+ H^x \,s\, \bar b)$ as a function of $\lambda$
at $\sqrt{s} = 500$ GeV, 1000 GeV, 1500 GeV,
for $h^0$ (left), $H^0$ (right).}
\label{resonancelam}
\end{center}\vspace*{-0.5cm}
\end{figure}

In Fig.~\ref{resonance} we show the behaviour of the cross section
with the center of mass energy,
based on the dominant contributions with
the $A^0$ intermediate states, Fig.~\ref{diagsresonancia}{(b)},
for both $h^0$ (left panel) and $H^0$ (right panel). 
Since the results for the contributions with $H^0$ intermediate states,
Fig.~\ref{diagsresonancia}{(c)}, are very similar, we do not include
explicit numerical results for this case. The behaviour of the cross
section with both the center of mass energy and the flavour parameter 
$\lambda$ for the case of $A^0$ intermediate states are straightforwardly
applied to the $H^0$ case.
If not stated differently, the value of $\lambda$ 
is fixed to be $\lambda=0.6$.
The variation of  the cross section with $\lambda$
is contained in Fig.~\ref{resonancelam}, 
shown for three different values of the energy, 
$\sqrt{s} = 500\,, \, 1000\,, \, 1500$ GeV.
The effects increase with $\lambda$ for both kind of Higgs bosons 
and vanish at $\lambda=0$, as to be expected. 
In the $H^0$ case, due to kinematical reasons, the cross section at
$\sqrt{s} = 500$ GeV is smaller than at 1 TeV, 
leading to an inversion of the order of the curves. In particular
for the light Higgs boson $h^0$, the cross section remains rather small, 
although it is effectively for a two-particle process with subsequent decay.
The resonating behaviour is largely canceled by either a small
$Zh^0A^0$ coupling or a tiny branching ratio of the $Z$ boson.   
Therefore, also non-resonant contributions have to be considered for
a reliable estimate of the total production rate.  

\subsection{Non-resonant contributions}
\label{chap.phenomenology}

\begin{figure}[hbt!]
\begin{center}\hspace*{-0.8cm}
\epsfig{file=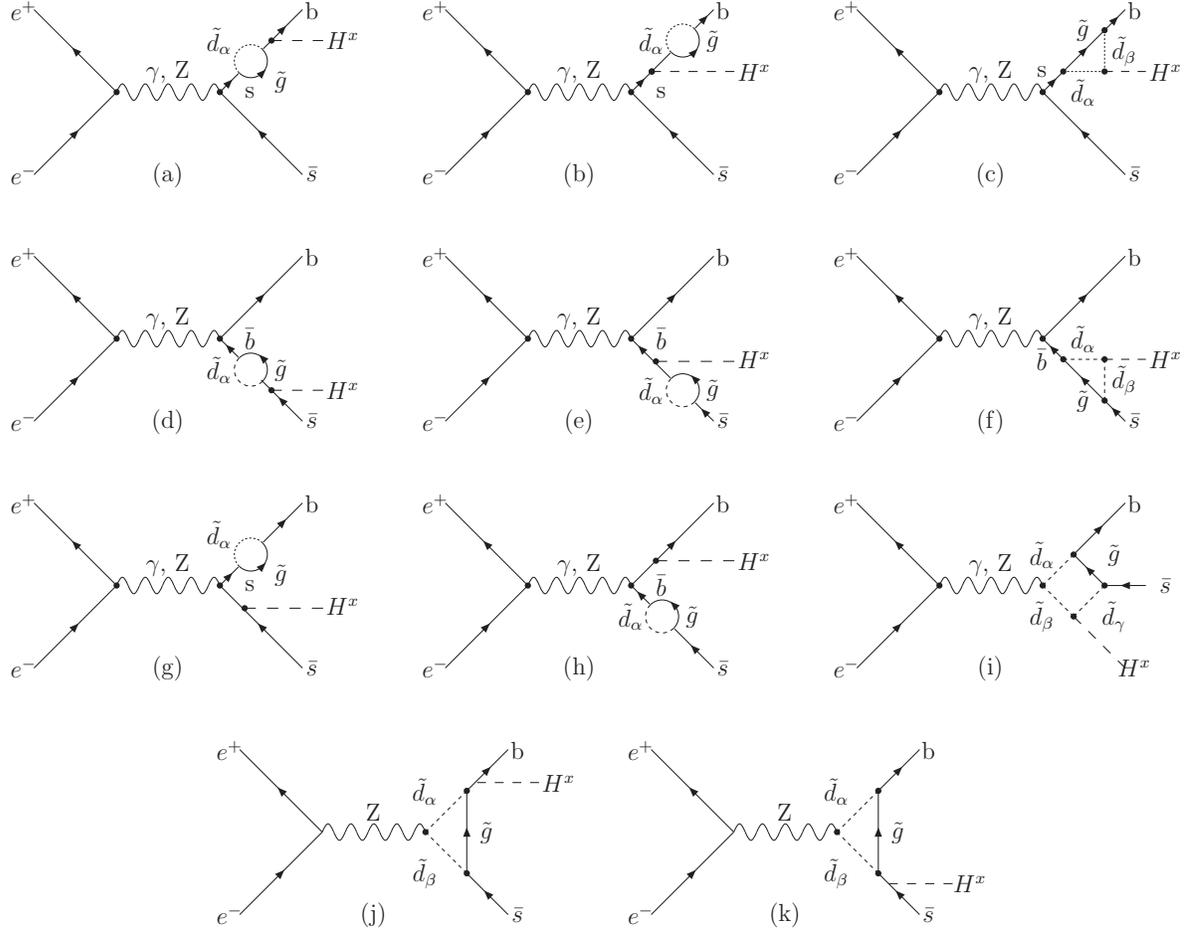,width=14cm}\vspace*{-0.2cm}
\caption{Generic diagrams for the squark-gluino one-loop contributions to
$e^{+} e^{-} \rightarrow 
H^x \, b\, \bar s+H^x \, s \,\bar b$ $(H^x \equiv h^0\,,H^0\,, A^0)$.}
\label{diagrams}\end{center}
\end{figure}
Besides the two-particle-like processes discussed in the previous subsection,
there are non-resonating contributions for $H^x b \bar{s}$ production,
which are genuine $2 \rightarrow 3$ processes.
The set of diagrams under consideration is shown in Fig.~\ref{diagrams}. 
There are three different types of relevant topologies:
self-energy, triangle,  and box diagrams. 
For the self-energy diagrams one has to distinguish between 
two different cases of Higgs radiation, one with
radiation from $b$-lines and the other one 
with radiation from $s$-lines. 
Because of the small Yukawa couplings for the $s$ case,
only Higgs radiation from $b$-quarks will be of significance. 

Here we outline the size of the contributions 
of each class of diagrams to the cross section for illustrational purposes,
based on the same set of model parameters as in 
subsection~\ref{chap.resonances}.
The non-resonant $e^{+} e^{-} \rightarrow Z \rightarrow b s$ 
contributions with Higgs radiation from $b,s$-lines 
(Fig.~\ref{diagrams}{(j,k)}) are suppressed with respect to the other 
contributions, being of the order of
$10^{-12} - 10^{-14}$~pb, therefore they are not included in the analysis.

\psfrag{tot}{{\footnotesize{${\mbox{total}}$}}}
\psfrag{main}{{\footnotesize{${\mbox{main}}$}}}
\psfrag{boxes}{{\footnotesize{${\mbox{boxes}}$}}}
\psfrag{srad}{{\footnotesize{${\mbox{s-rad.}}$}}}
\begin{figure}[t]\begin{center}\vspace*{-0.5cm}
\hspace*{-0.4cm}
\raisebox{-0.2cm}{\epsfig{figure=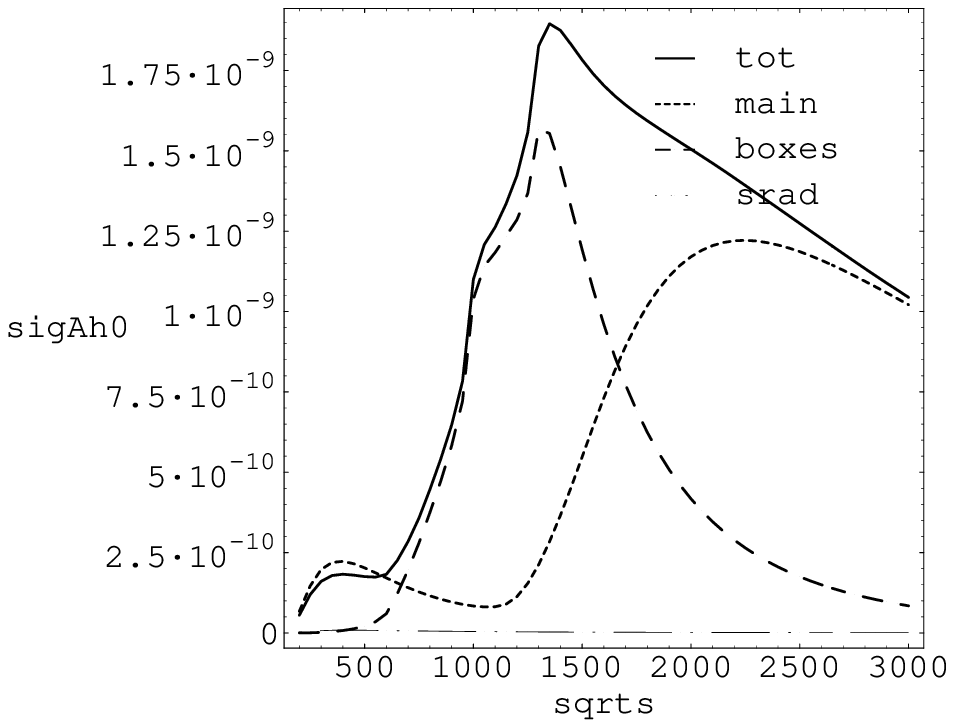,scale=0.82}}
\hspace*{-0.4cm}
\epsfig{figure=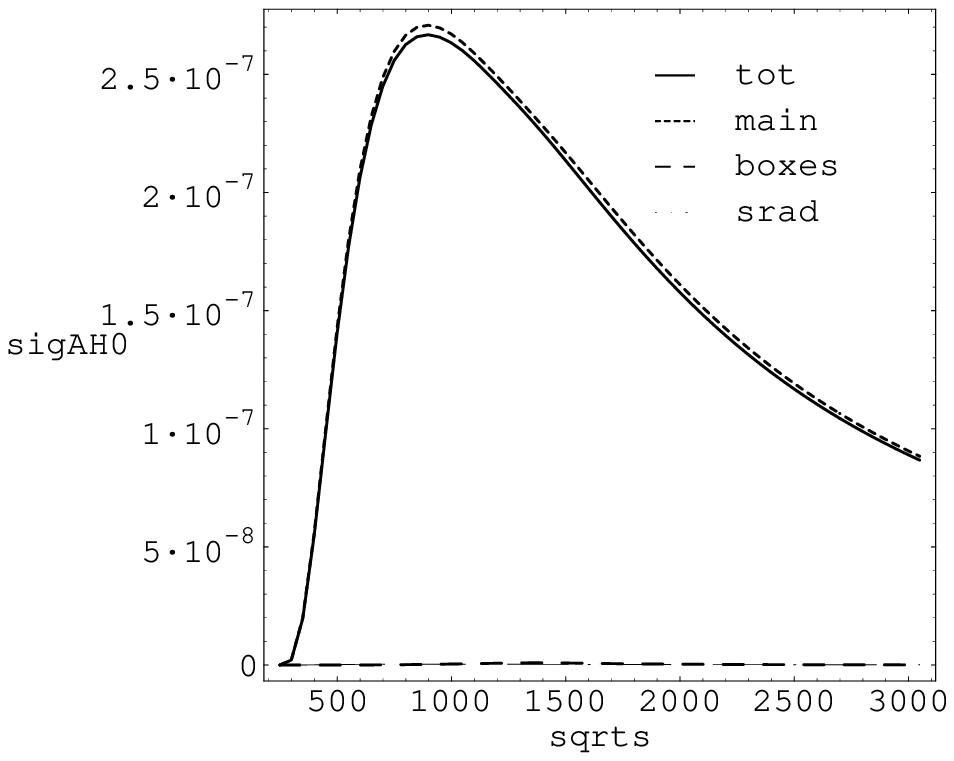,scale=0.78}
\vspace{-1.2cm}
\caption{One-loop contributions to $\sigma(e^{+} e^{-} \rightarrow 
H^x \, b\, \bar s+ H^x \,s\, \bar b)$ as a function of $\sqrt{s}$ (GeV).
Total cross sections, {\it main} and {\it box} contributions, and
corrections with Higgs radiation off the $s$-lines 
($h^0 \equiv$ left panel, $H^0 \equiv$ right panel). Here $\lambda=0.6$.}
\label{contrib_diags}\end{center}\vspace*{-0.4cm}
\end{figure}
\psfrag{sih0}{\rotatebox{90}{$\sigma \left[ \rm{pb} \right]$}}
\psfrag{siH0}{\rotatebox{90}{$\sigma \left[ \rm{pb} \right]$}}
\begin{figure}[htb!]
\begin{center}
\hspace*{-1cm}
\epsfig{file=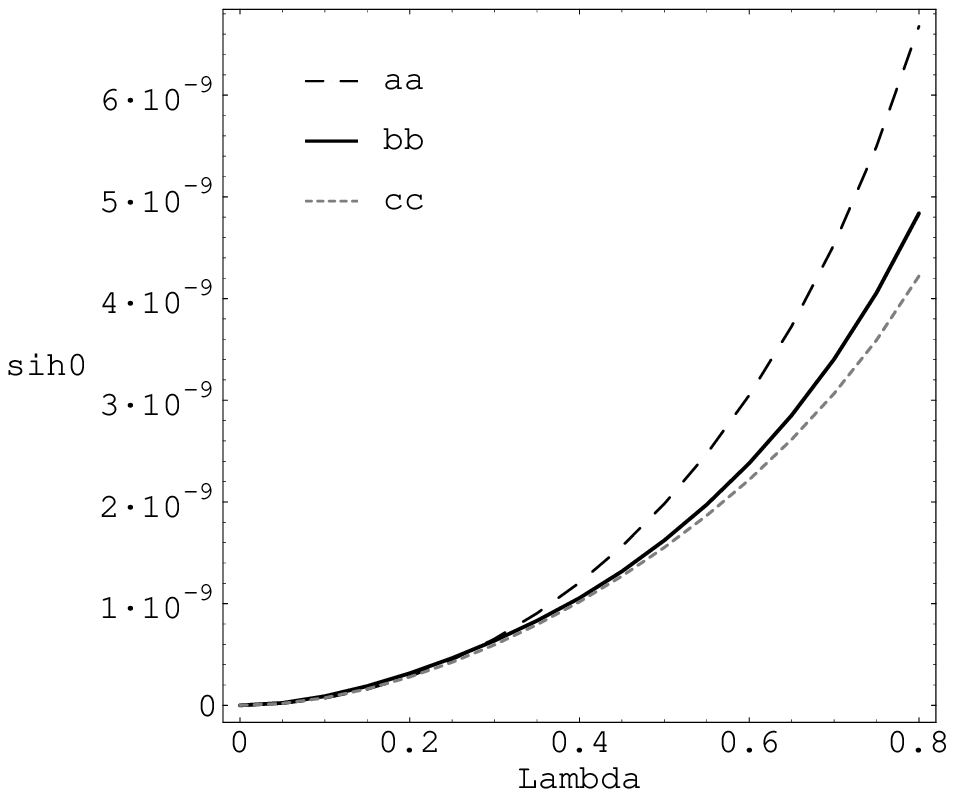,scale=0.75}
\epsfig{file=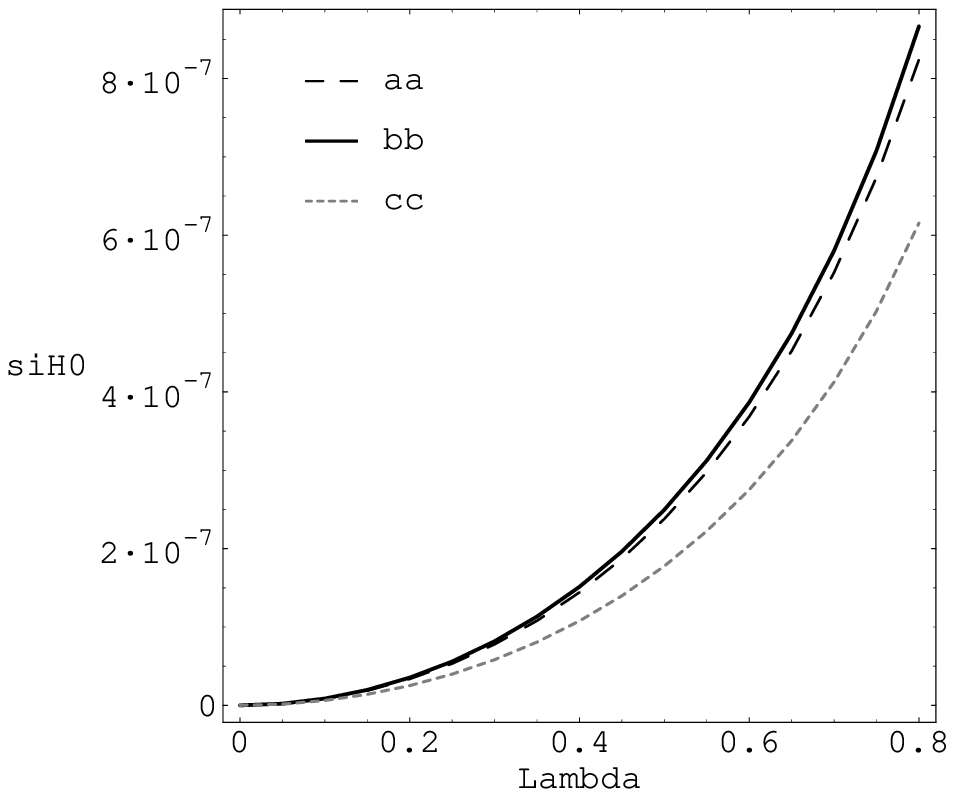,scale=0.75}\vspace*{-1cm}
\caption{$e^{+} e^{-} \rightarrow 
H^x \to b \bar s+ s \bar b$  cross section as a function of $\lambda$ at 
$\sqrt{s} = 500\,,\, 1000\,,\,1500$ GeV.}
\label{hbs_lambda}
\end{center}\vspace*{-0.4cm}
\end{figure}
In Fig.~\ref{contrib_diags} the one-loop contributions from the 
diagrams in Fig.~\ref{diagrams} are depicted as a function of 
the center of mass energy $\sqrt{s}$. Since the
results for $\sigma (e^{+} e^{-} \rightarrow     
H^0 \,b\, \bar s+ H^0 \,s\, \bar b)$ and $\sigma (e^{+} e^{-} \rightarrow
A^0\, b \,\bar s+A^0\, s\, \bar b)$ are very similar,
we do not include explicit numerical 
results for $A^0$ production. The discussion for the $H^0$ case can 
straightforwardly be applied to the $A^0$ case along this paper.  
Besides the total cross sections (solid line)
we list in addition the contributions 
from three different topologies independently: 
{\it main} contributions (short-dashed line), 
defined as the ones resulting from triangle diagrams and from
self-energy diagrams with Higgs radiation off the $b$-lines,
{\it box} contributions (long-dashed line), and
the {\it s-rad} contributions, defined as the self-energy contributions
with Higgs radiation off the $s$-lines. 
These last contributions are  
presented here for illustrational purposes, 
they are of the order of roughly $10^{-11}$ pb for both Higgs bosons.  
In the case of the heavy Higgs bosons, the box
contributions are smaller than in the $h^{0}$ case, being of the order
of $10^{-10} - 10^{-11}$~pb, therefore they appear together with the
contributions with Higgs radiation on the $s$-lines near to zero.
For the case of $h^0$ production, 
there are several peaks corresponding to squark-mass
thresholds in the loop integrals. 
These peaks are less distinctive in the $H^0$ case because they appear
only in the box contributions, which are smaller for $H^0$. 

One can see from Fig.~\ref{contrib_diags} 
that for both Higgs bosons the non-resonant 
contributions are roughly two orders of
magnitude smaller than the resonant contributions analyzed before. 
The dependence of the cross sections on
the mixing parameter $\lambda$ 
is illustrated in Fig.~\ref{hbs_lambda} at 1 TeV.
As before, the cross sections increase with
$\lambda$, being exactly zero for $\lambda=0$.

\section{Dependence on the MSSM parameters.}
\label{chap.numerical}

\psfrag{TB}{$\tan \beta$}
\begin{figure}[t!]
\begin{center}
\raisebox{0.4cm}{\epsfig{figure=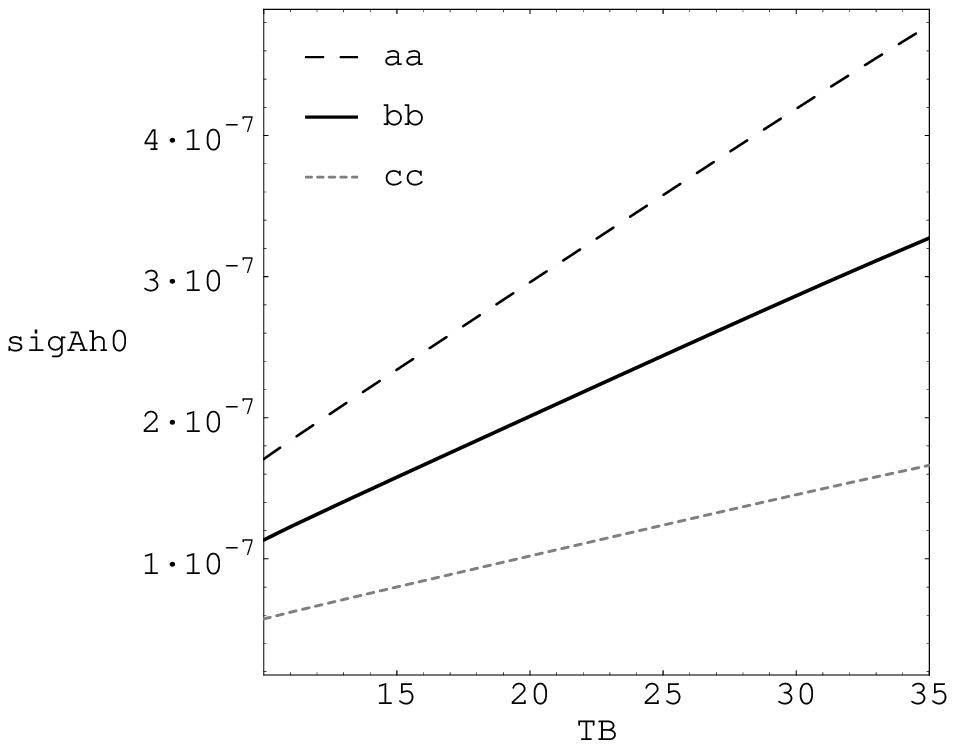,scale=0.76}}\hspace*{-0.4cm}
\epsfig{figure=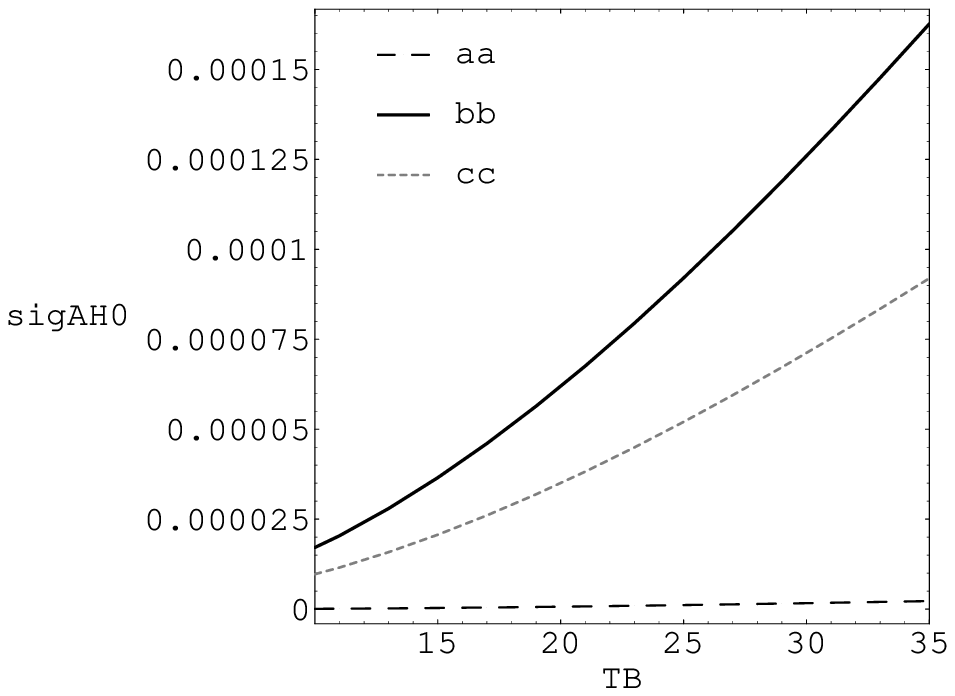,scale=0.83}\\
\vspace*{-1.5cm}(a)\vspace*{-0.3cm}\\
\raisebox{0.15cm}{\epsfig{file=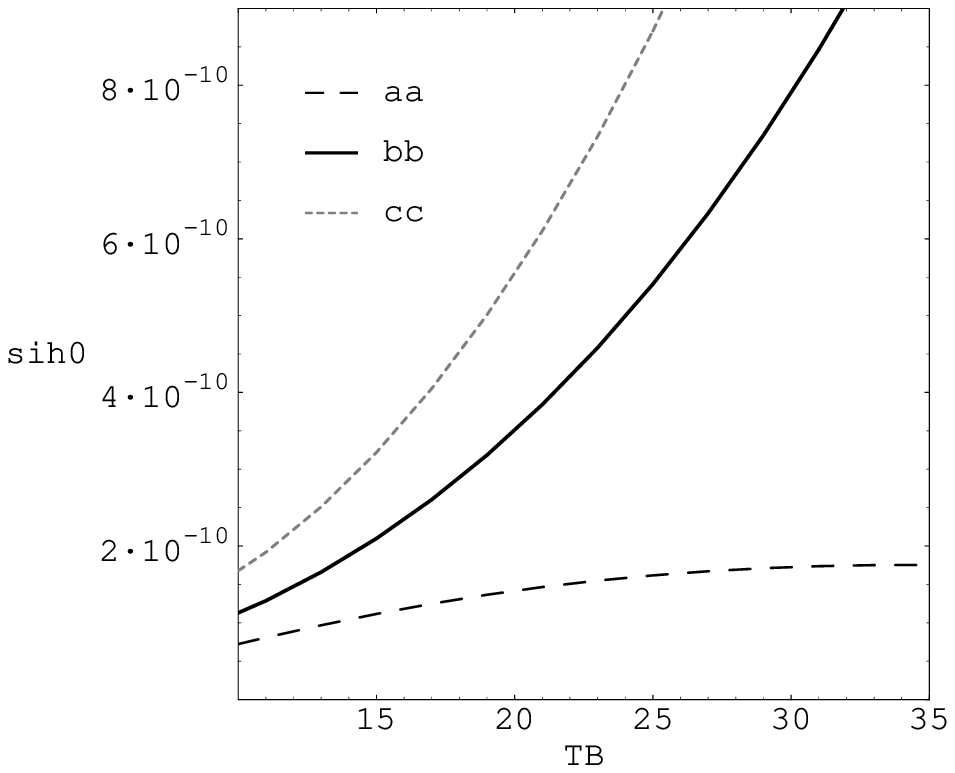,scale=0.75}}\hspace*{-0.1cm}
\epsfig{file=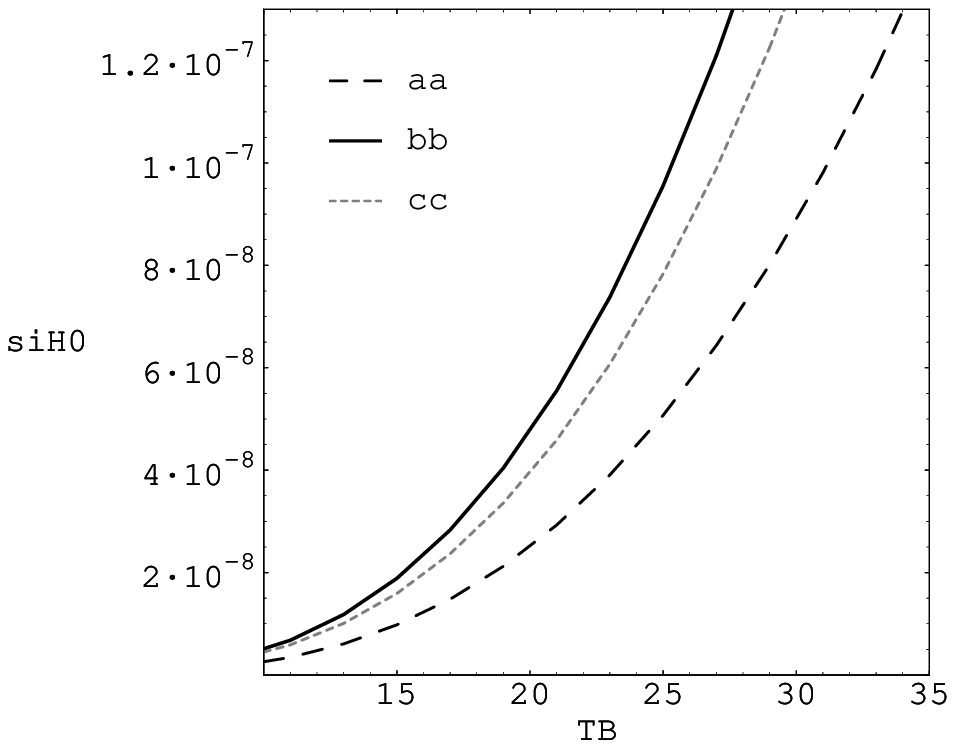,scale=0.78}\\
\vspace*{-1.1cm}(b)
\caption{Cross section as a function
of $\tan \beta$, at $\sqrt{s} = 500\,, \, 1000\,, \, 1500$ GeV for:
{\bf{(a)}} $e^{+} e^{-} \rightarrow A^0 H^x \rightarrow 
H^x \, b\, \bar s+H^x \,  s\, \bar b$,
{\bf{(b)}} non-resonant
$e^{+} e^{-} \rightarrow H^x \, b\, \bar s+H^x \,  s\, \bar b$.}
\label{hbs_tbeta_res}
\end{center}\vspace{-0.5cm}
\end{figure}
The SUSY parameter set needed to determine the input for 
$e^{+} e^{-} \rightarrow 
H^x \, b\, \bar s + H^x\, s\, \bar b\,$ $(H^x \equiv h^0, H^0)$ 
consists of the quantities
$m_A$, $\tan \beta$, $\mu$, $M_{\tilde g}$, $M_0$, $A$, where $M_0$ is 
a common value for the soft-breaking squark mass parameters, 
$M_0 = M_{\tilde Q,q} = M_{\tilde U,(c,t)} = M_{\tilde D,(s,b)}$,
and $A$ denotes the trilinear parameters which are chosen to be equal,
$A_t=A_b=A_c=A_s=A$; $M_{\tilde g}$ is the mass of the gluino.
These six parameters will be varied over a broad range, subject to the 
 requirements that all 
 the squark masses be heavier than 150 GeV~\cite{PDB} and 
$M_{h_{0}}> 114.4$ GeV~\cite{search}. 
Similarly, in view of the present experimental bounds on the chargino
 mass~\cite{PDB}, we consider only $|\mu|$ values above $90$ GeV.   
The flavour mixing parameter, $\lambda$, is constrained by
the lower squark-mass bounds.
For those parameters that are kept fixed in the various figures
the following default set has been chosen,
\bea
\label{eq:numparameters}
\mu &=& 1000 \, {\rm GeV}\,, M_0=500 \,{\rm GeV}\,,  A = 800 \,{\rm GeV}\,,\nonumber\\
m_A &=& 250 \, {\rm GeV}\,, M_{\tilde g}=800 \,{\rm GeV}\,, \tan \beta = 35 \,,
\eea
which is in accordance with
experimental bounds for the decay $b\rightarrow s
\gamma$~\cite{boundsbsg}, as checked  
with the help of the code {\it micrOMEGAs}~\cite{Omega}, based on
minimal flavour violation calculations at leading
order~\cite{leadingbsg} and some
contributions beyond leading order that are important for high
values of $\tan \beta$~\cite{gambinobsg}, as well as in accordance with 
the result for the muonic $(g-2)$. 
The FCNC effects in the rare decay 
$b\rightarrow s \gamma$ may impose severe constraints on MSSM parameters
that reduce the maximally allowed value for $BR (h^0  
\rightarrow  b\, \bar s)$~\cite{sola}. 
Furthermore, starting from~\eqref{eq:numparameters}, 
$\lambda > 0.8$ implies unallowed values for the squark
masses, i.e. $M_{\tilde q} < 150$ GeV. 
If not stated differently, $\lambda=0.6$ is a default value.
\psfrag{tanb}{{$\,\tan \beta$}}
\psfrag{TB}{{$\,\tan \beta$}}

In the following we explore the dependence of the cross
section on the MSSM parameters for both resonant and non-resonant
diagrams, illustrated in 
Figs.~\ref{hbs_tbeta_res} to~Figs.~\ref{hbs_M0_res};
therein, the upper
panels correspond to the resonant contributions and the lower panels show the
non-resonant contributions. 

\psfrag{MUE}{{$\,\mu$}}
\psfrag{sigZAh0}{\rotatebox{90}{$\sigma \left[ \rm{pb} \right]$}}
\psfrag{sigZH0}{\rotatebox{90}{$\sigma \left[ \rm{pb} \right]$}}
\begin{figure}[t!]
\begin{center}
\raisebox{0.1cm}{\epsfig{figure=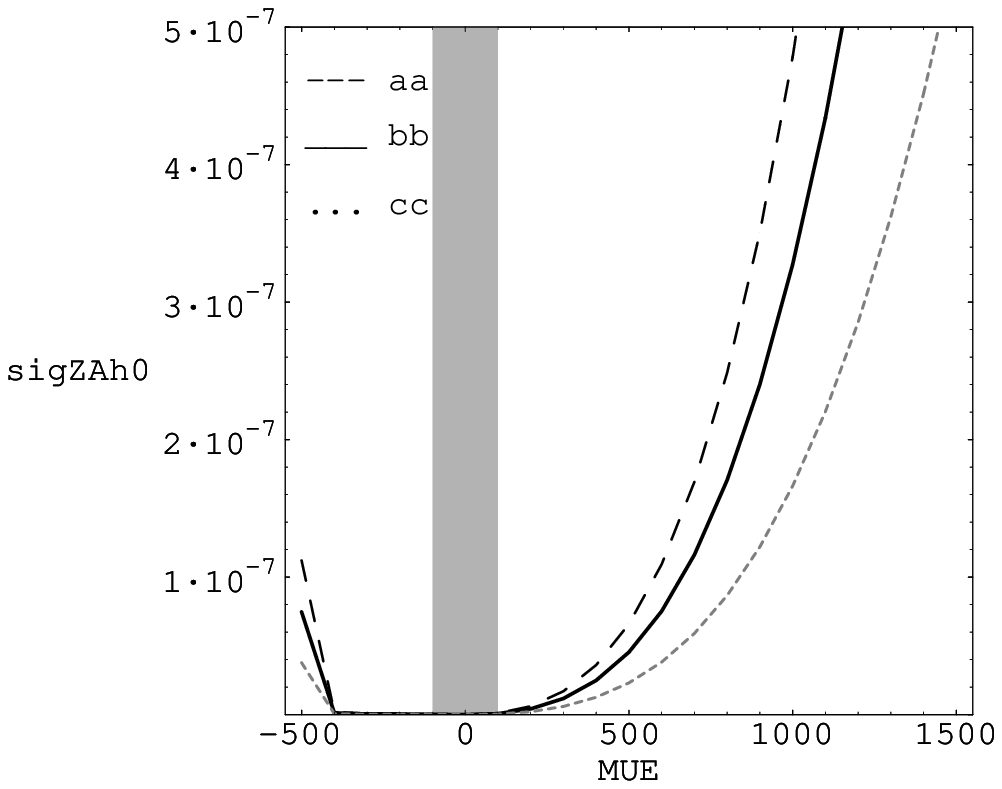,scale=0.75}}\hspace*{-0.1cm}
\epsfig{figure=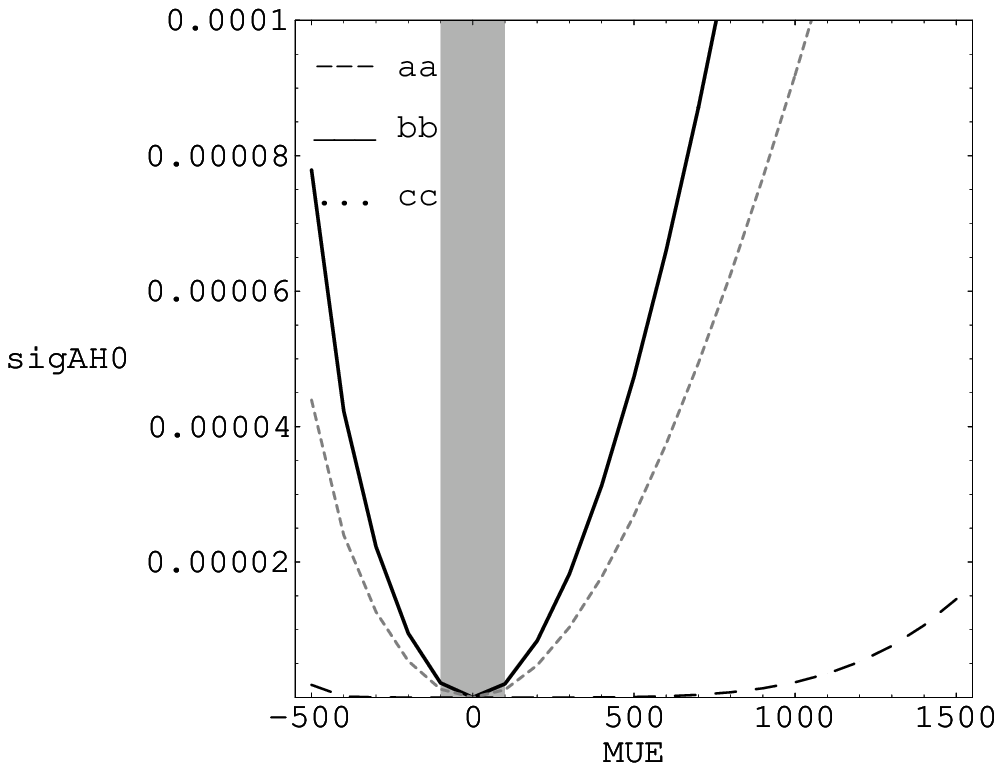,scale=0.76}\\
\vspace*{-1cm}(a)\vspace*{-0.3cm}\\
\hspace*{-0.3cm}\epsfig{file=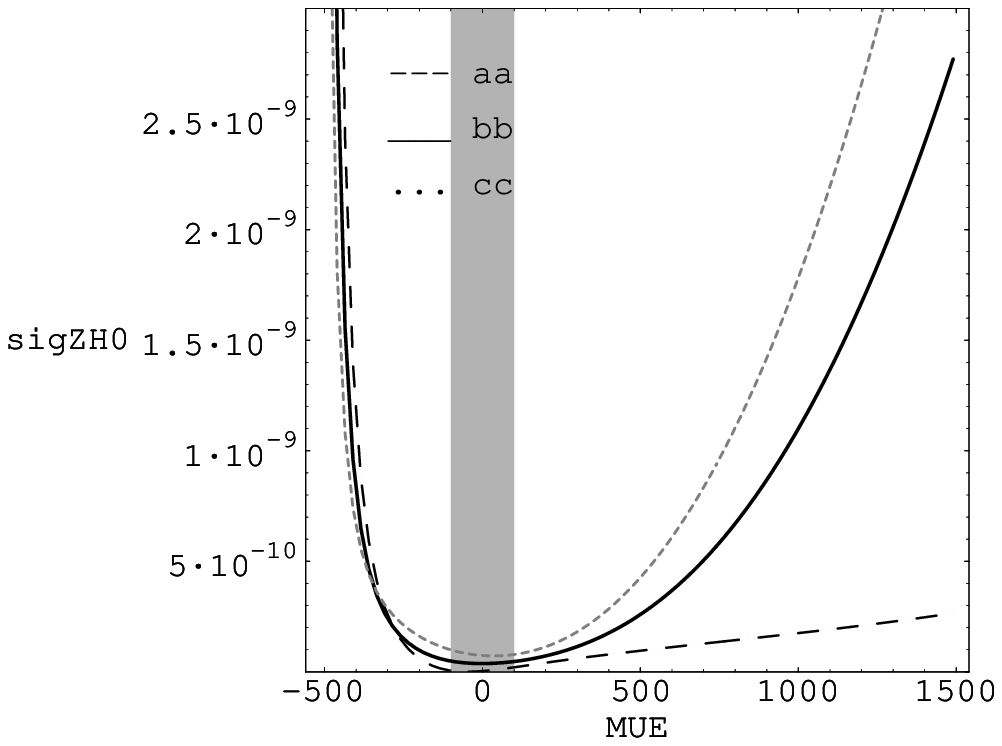,scale=0.79}\hspace*{-0.1cm}
\raisebox{0.35cm}{\epsfig{file=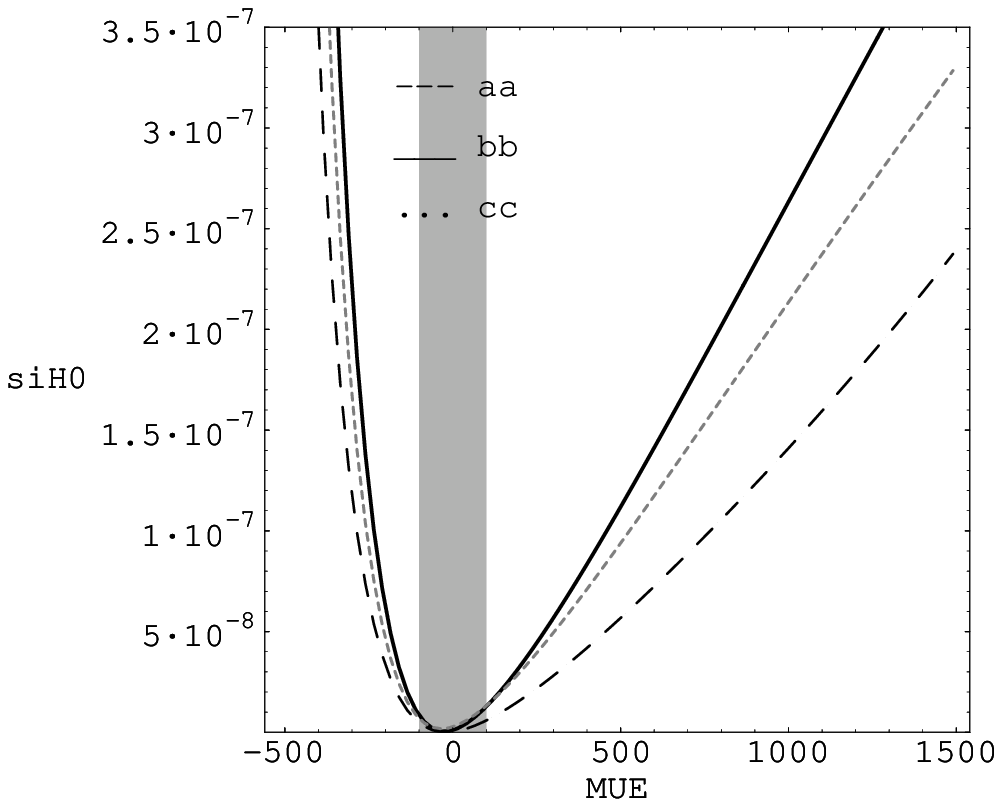,scale=0.74}}\vspace*{-0.3cm}\\
\vspace*{-0.8cm}(b)
\caption{Cross section as a function as a function
of $\mu$ (GeV) at $\sqrt{s} = 500\,,\, 1000\,,\, 1500$ GeV for:
{\bf{(a)}} $e^{+} e^{-} \rightarrow A^0 H^x \rightarrow 
H^x \, b\, \bar s+H^x \,  s\, \bar b$, 
{\bf{(b)}} non-resonant
$e^{+} e^{-} \rightarrow H^x \, b\, \bar s+H^x \,  s\, \bar b$.}
\label{depenmu_res}
\end{center}\vspace{-0.5cm}
\end{figure}
First, we discuss the
dependence on $\tan \beta$, the parameter that appears inside 
the couplings as well as in the squark-mass matrices
and thus determines the squark-mass splitting
and the b-mass corrections. 
Fig.~\ref{hbs_tbeta_res} contains
the cross section $\sigma (e^{+} e^{-} \rightarrow 
H^x\, b\, \bar s+H^x\, s\, \bar b)$ 
versus $\tan \beta$ for three
values of the center mass energy, $\sqrt{s} = 500\,,\, 1000\,,\, 1500$ GeV. 
In all cases, the cross sections increasing with $\tan\beta$.
For the $h^0$ boson (left panel) 
the  increase with $\tan \beta$ is slower due to 
a superposition of two different  
effects in the involved subprocesses: the reinforcement
in the decay subprocess from the coupling enhancement is 
partly compensated by decreasing in $\cba$ in the production cross section 
of the $2\rightarrow 2$ process, which causes the almost linear shape
of the $\tan \beta$ dependence. 
Values of $\tan \beta < 10$ are not included in these plots
because they imply Higgs
boson mass values lower than $114.4$ GeV. 

\psfrag{MGl}{{$M_{\tilde g}$}}
\psfrag{@GeVDgu}{{$[GeV]$}}
\psfrag{sigh0}{\rotatebox{90}{$\sigma \left[ \rm{pb} \right]$}}
\psfrag{sigH0}{\rotatebox{90}{$\sigma \left[ \rm{pb} \right]$}}
\begin{figure}[tb!]
\begin{center}
\raisebox{0.2cm}{\epsfig{figure=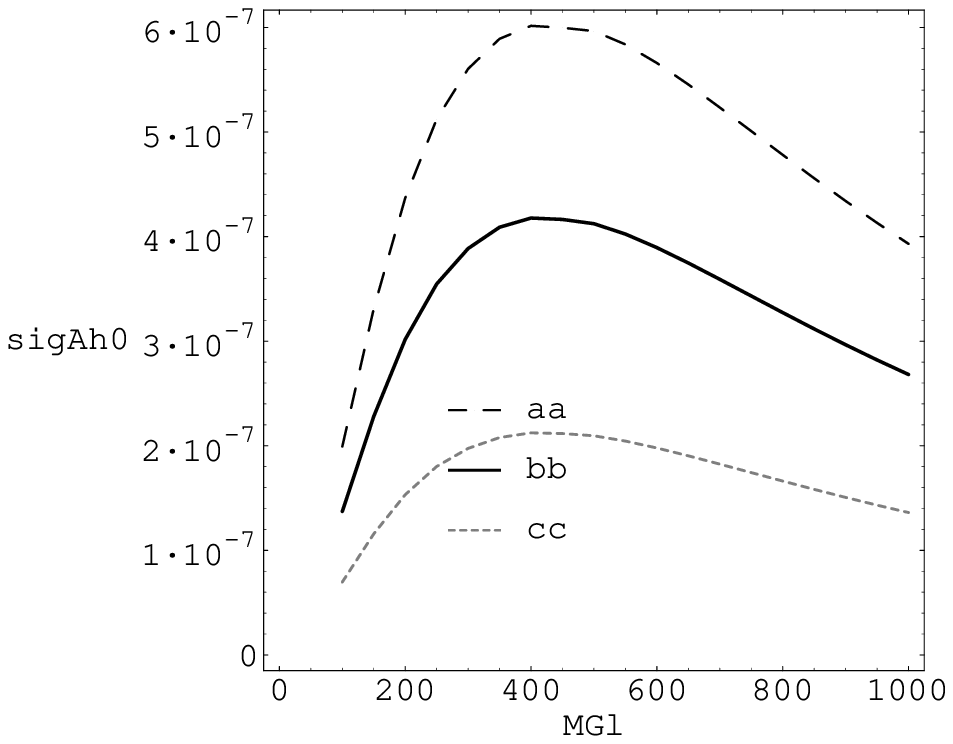,scale=0.77}}\hspace{-0.3cm}
\epsfig{figure=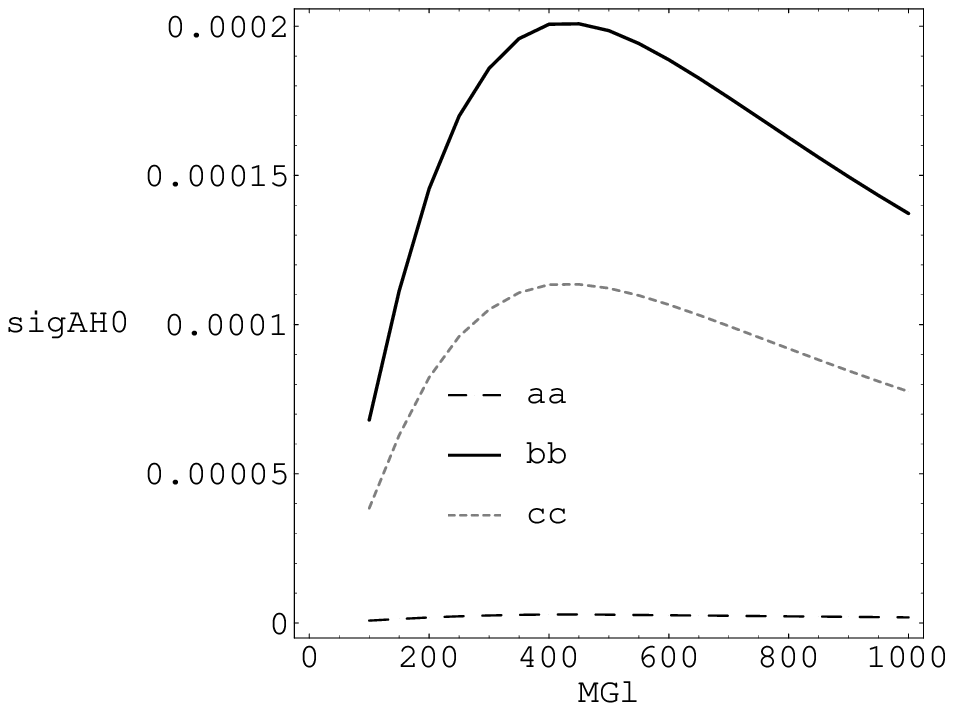,scale=0.81}\\
\vspace*{-1.4cm}(a)\vspace*{-0.2cm}\\
\raisebox{0.25cm}{\epsfig{file=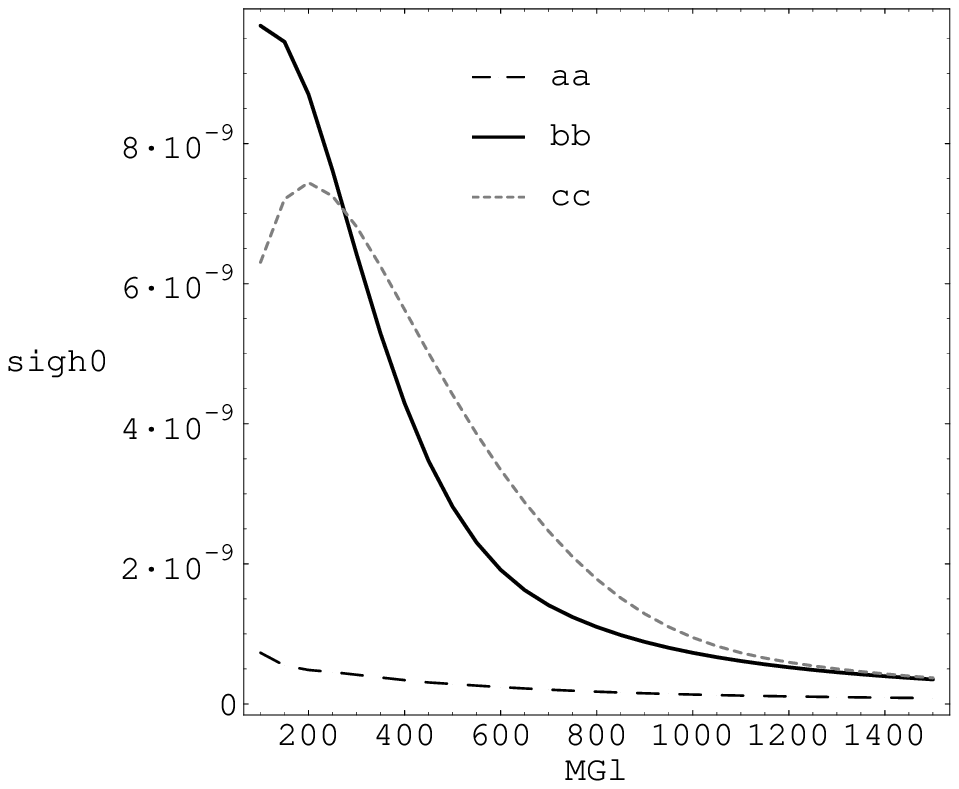,scale=0.73}}\hspace*{-0.3cm}
\epsfig{file=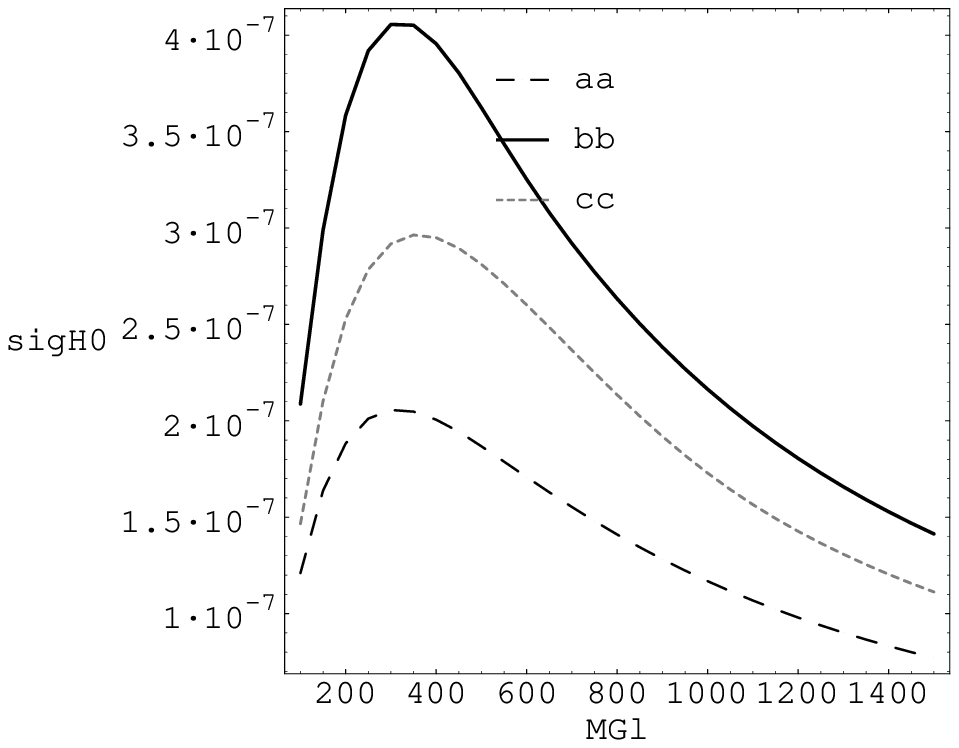,scale=0.77}\\
\vspace*{-1.1cm}(b)
\caption{Cross section as a function
of $M_{\tilde g}$ (GeV) at $\sqrt{s} = 500\,,\, 1000\,,\, 1500$ GeV for:
{\bf{(a)}} $e^{+} e^{-} \rightarrow A^0 H^x \rightarrow 
H^x \, b\, \bar s+H^x \,  s\, \bar b$, 
{\bf{(b)}} non-resonant
$e^{+} e^{-} \rightarrow H^x \, b\, \bar s+H^x \,  s\, \bar b$.}
\label{hbs_Mg_res}
\end{center}\vspace{-0.5cm}
\end{figure}
The dependence on the $\mu$ parameter for the $h^0$ (left) 
and $H^0$ case (right) is shown in Fig.~\ref{depenmu_res}.  
The shaded regions in these figures correspond to the regions excluded
by the LEP bounds on the chargino mass, $|\mu| \leq 90$ GeV.
One can see that in all cases the results for the cross sections
are not symmetric under $\mu \to -\mu$. 
This is essentially due to the inclusion of the threshold corrections to
the bottom quark mass, where the sign of $\mu$ plays an important role. 
We found that the corrections grow with the $\mu$ parameter. 

\psfrag{MSusy}{$M_0$}
\begin{figure}[tbh!]
\begin{center}
\epsfig{figure=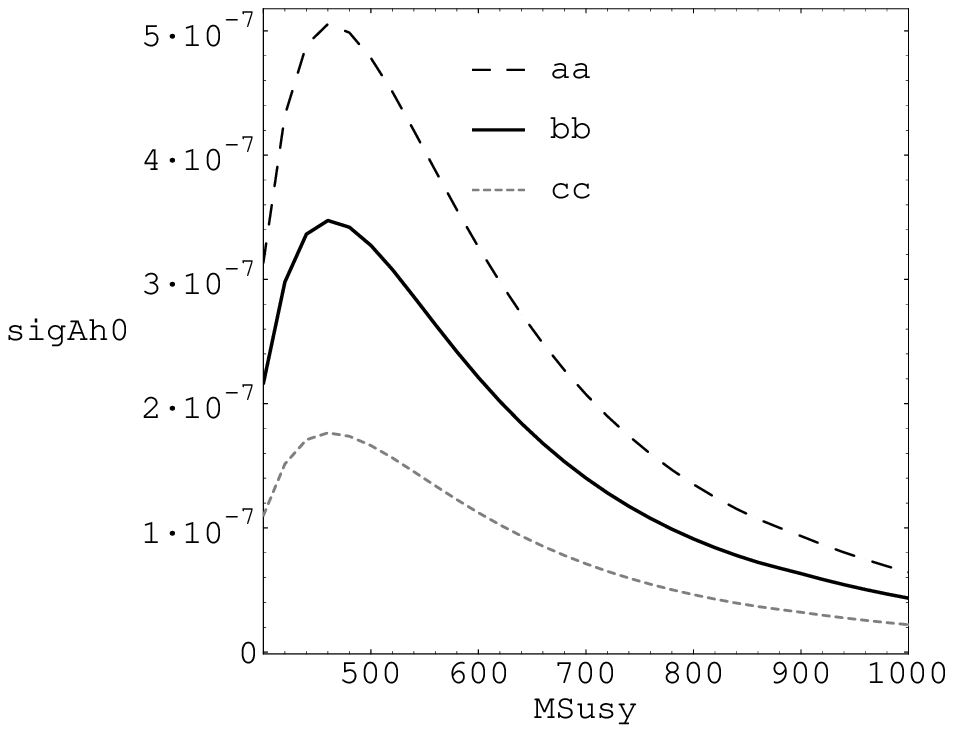,scale=0.80}\hspace{-0.2cm}
\epsfig{figure=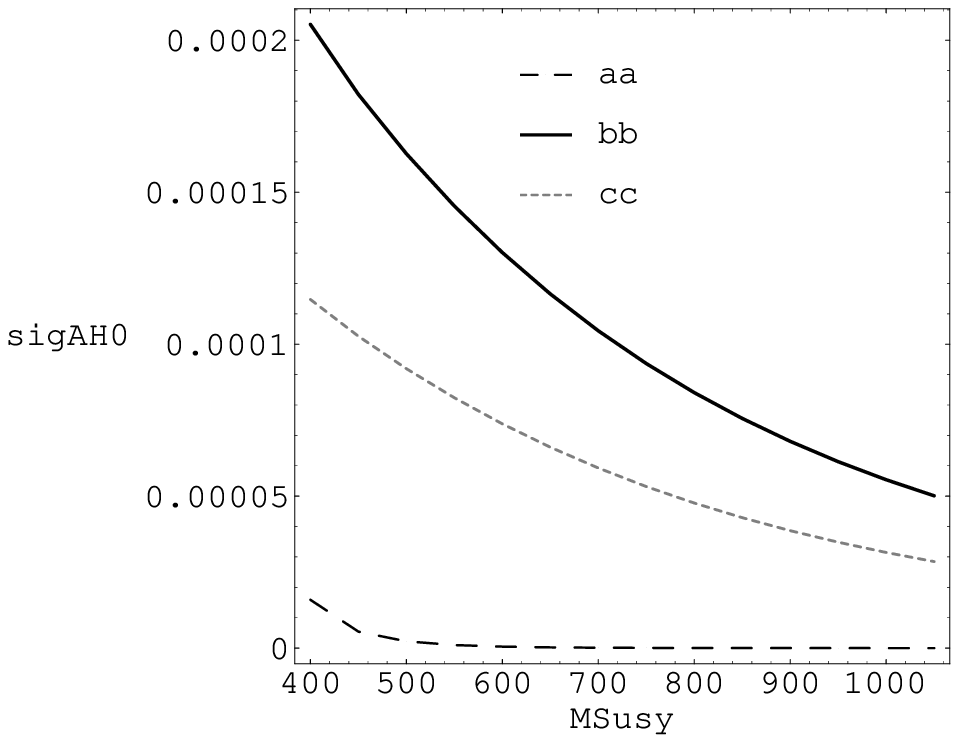,scale=0.78}\\
\vspace*{-1.1cm}(a)\vspace{-0.5cm}\\
\hspace*{-0.5cm}\raisebox{0.05cm}{\epsfig{file=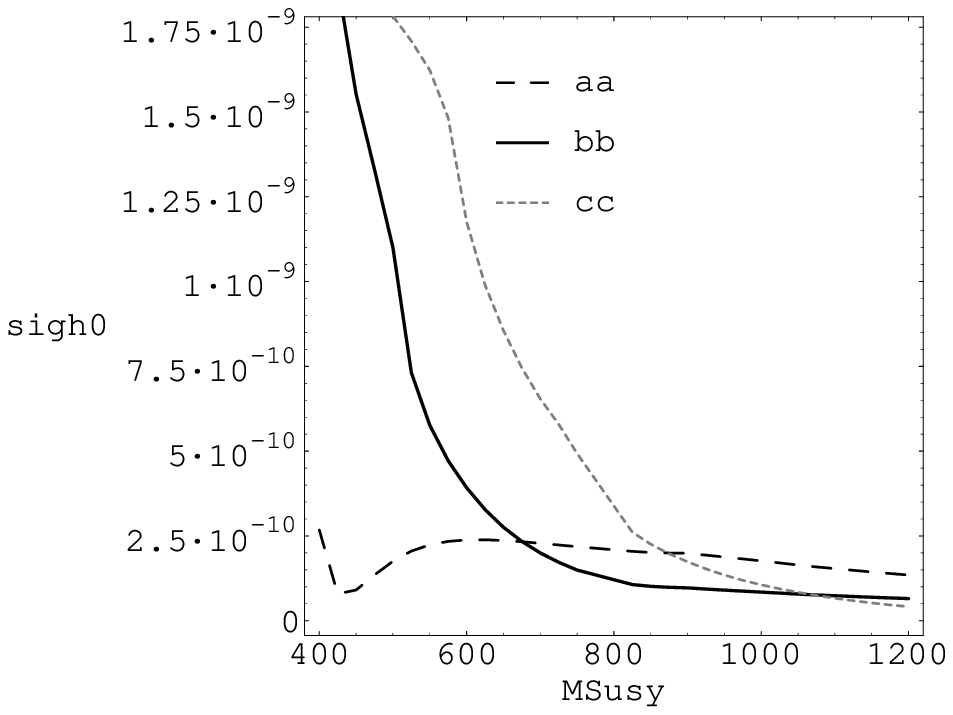,scale=0.83}}
\hspace*{-0.3cm}\raisebox{0.2cm}{\epsfig{file=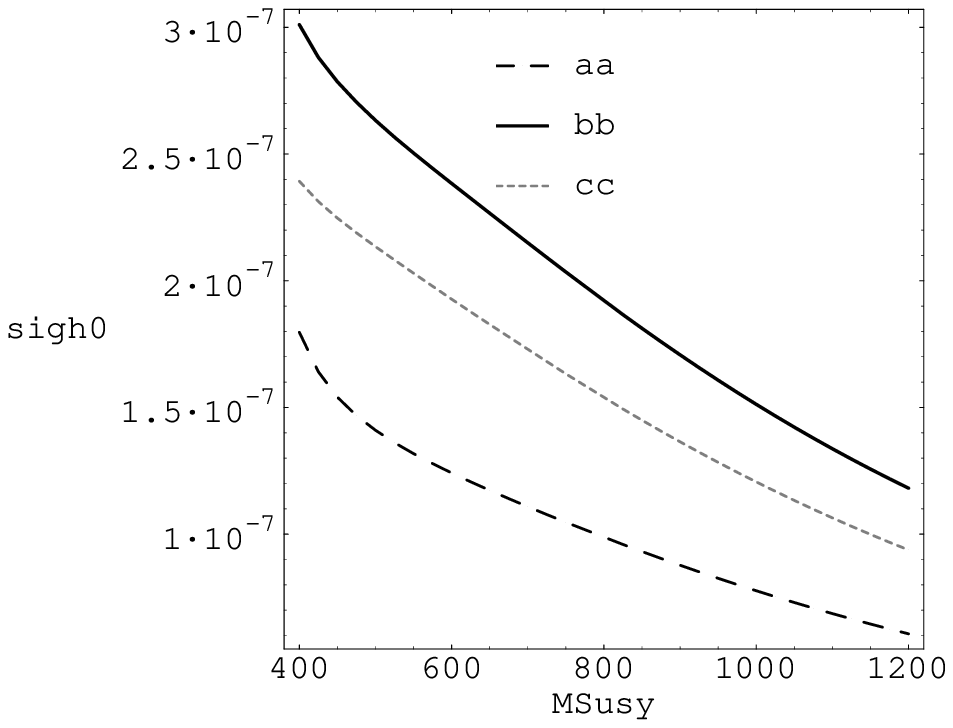,scale=0.80}}\\
\vspace*{-1.1cm}(b)
\caption{Cross section as a function
of $M_0$ (GeV) at $\sqrt{s} = 500\,,\, 1000\,,\, 1500$ GeV for:
{\bf{(a)}} $e^{+} e^{-} \rightarrow A^0 H^x \rightarrow 
H^x \, b\, \bar s+H^x \,  s\, \bar b$, 
{\bf{(b)}} non-resonant
$e^{+} e^{-} \rightarrow H^x \, b\, \bar s+H^x \,  s\, \bar b$.}
\label{hbs_M0_res}
\end{center}
\end{figure}
Next we study the behaviour with respect to the other MSSM parameters.
It turns out that the cross sections are
nearly independent of the common trilinear parameters $A$. 
In Fig.~\ref{hbs_Mg_res} we show the dependence 
on the gluino mass, $M_{\tilde g}$.
One  can see from this figure that for all cases the 
FCNC effects vanish for heavy gluinos, i.e.\ the gluino decouples 
when only the gluino mass is varied. 
It is also interesting to see that the maximum is reached not for 
the smallest values but for intermediate values of the gluino mass.
\psfrag{MSusy}{{$\,M_0$}}

The dependence of the cross sections 
on the common soft SUSY breaking squark mass parameter $M_{0}$ 
GeV is displayed in Fig.~\ref{hbs_M0_res}.
Regions below $M_0 = 400$ GeV are not dropped
because they would lead to excluded values for the down-squark masses. 
The FC cross sections decrease
with  $M_{0}$, showing the independent
decoupling behaviour of $M_{0}$. 

In summary, the study of the dependence on the MSSM parameters, 
has confirmed the maximum size of the cross sections of
the order of $10^{-7}$ pb for $h^0$ and
${\cal{O}} (10^{-4})$ pb for the $H^0$ case
and demonstrates the individual decoupling behaviour
of the virtual SUSY particles.

\section{Non-decoupling behaviour of heavy SUSY particles}
\label{chap.decoupling}

Here we analyze the non-decoupling behaviour of squarks and
gluinos in SUSY-QCD contributions to our FCNC $e^+ e^-$ processes
that result in a non-vanishing cross section also for very heavy
SUSY particles. Non-decoupling behaviour of
 SUSY particles also occurs in the SUSY-QCD
 and SUSY-EW contributions to FCNC Higgs boson decays~\cite{maria,florian}. 
\psfrag{Ms}{{$M_{S}$}}
\psfrag{M_{o}}{{$M_{S}$}}
\begin{figure}[tbh!]
\begin{center}\hspace*{-0.6cm}
\epsfig{figure=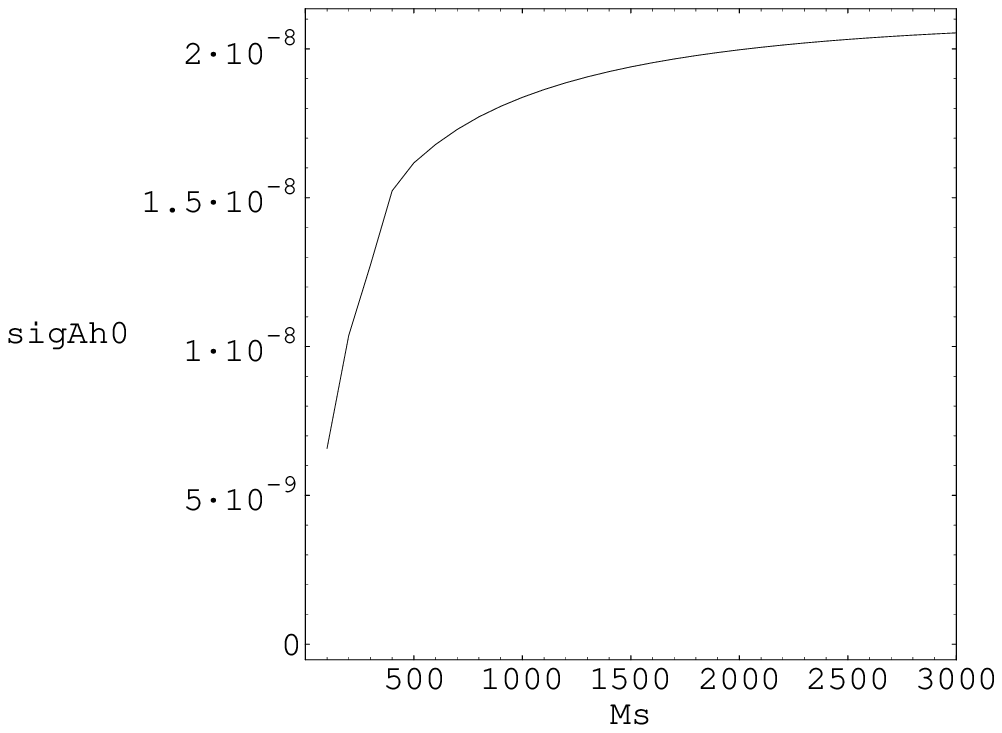,scale=0.80}\hspace*{-0.2cm}
\epsfig{figure=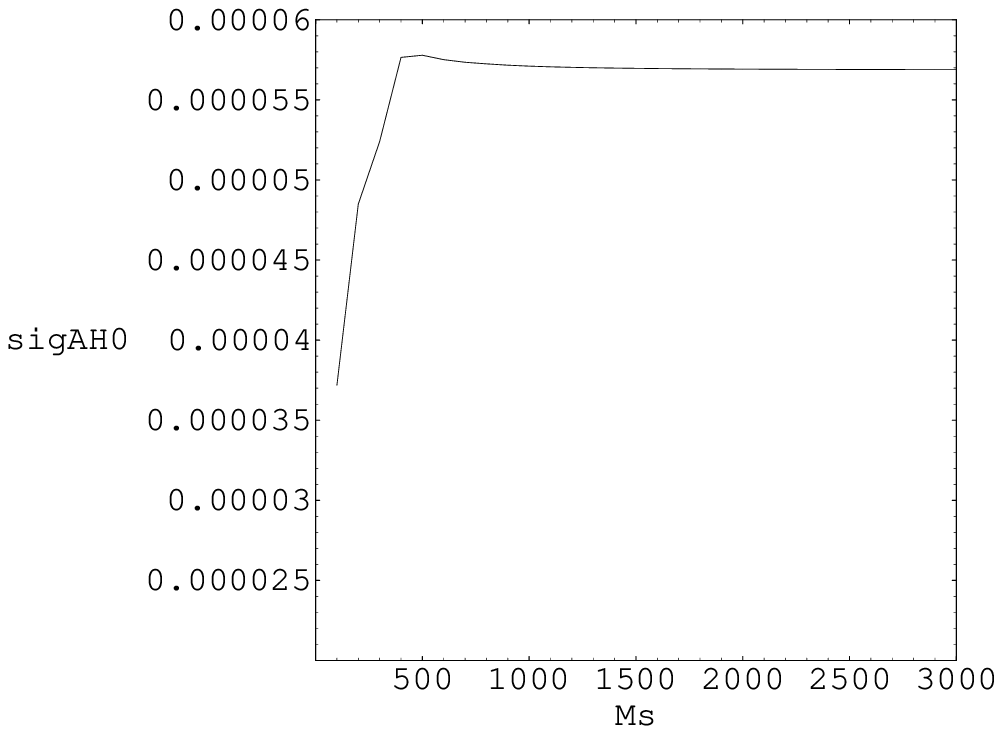,scale=0.80}\vspace*{0.3cm}\\
\epsfig{file=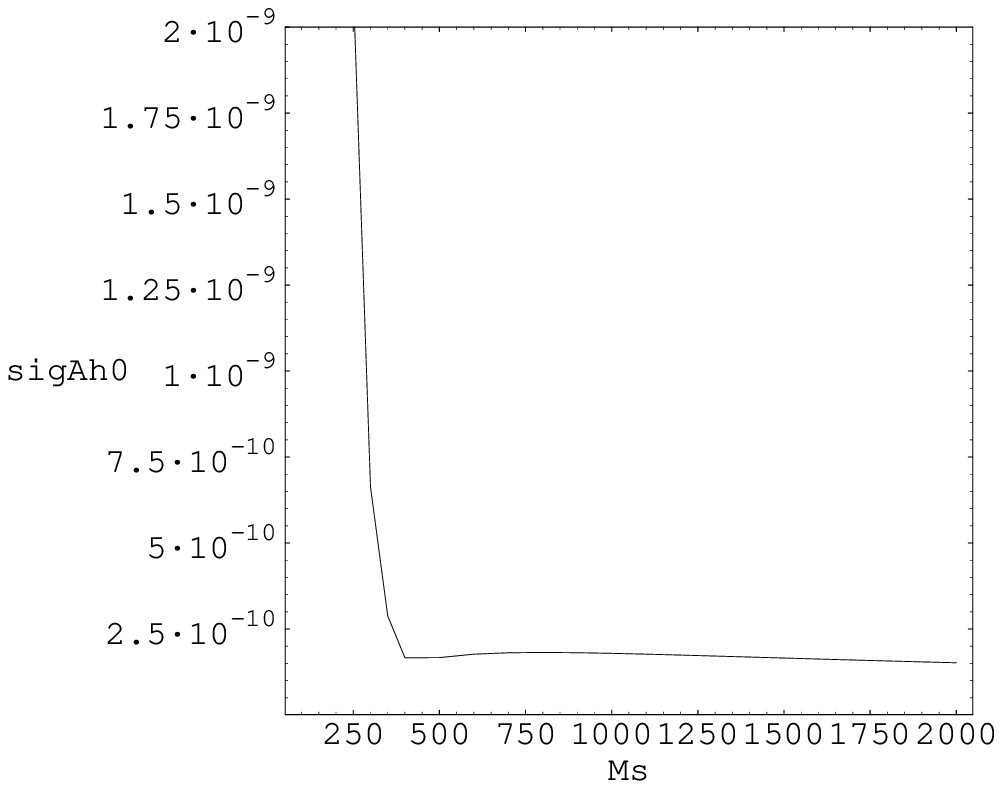,width=7.7cm}\hspace*{-0.1cm}
\epsfig{file=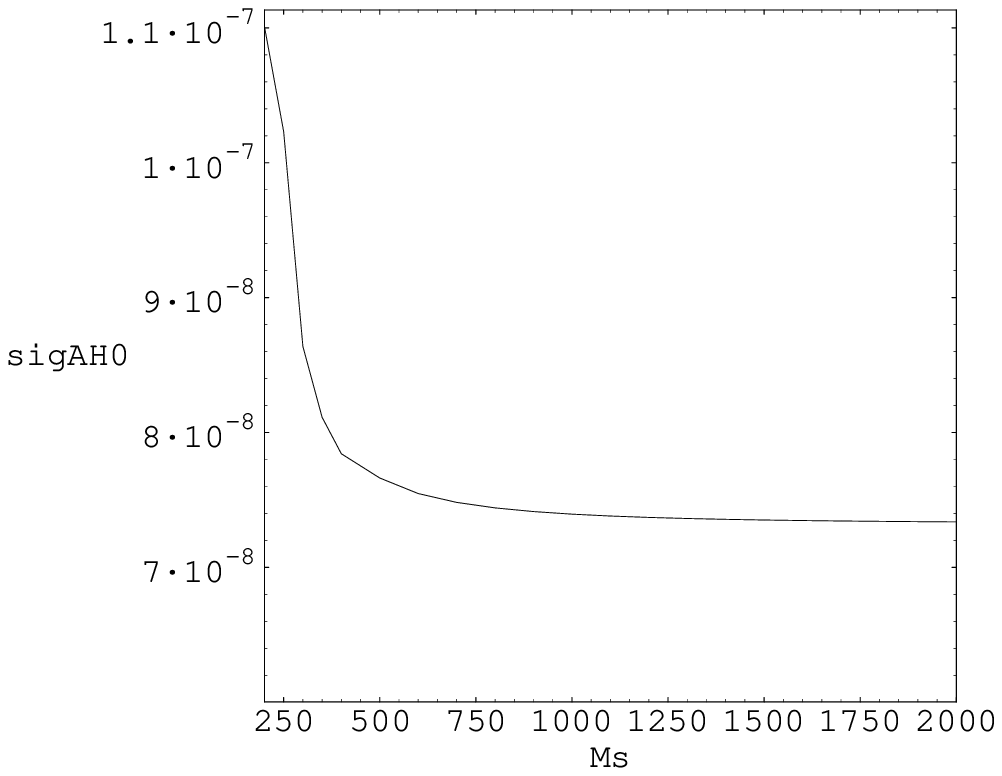,width=7.7cm}
\caption{Non-decoupling behaviour of the cross section 
from resonant contributions (upper panels), and of 
$\sigma(e^{+} e^{-} \rightarrow H^x \to b \bar s+ s \bar b)$ 
from non-resonant contributions (lower panels). $M_S$, in
GeV, is defined in~(\ref{eq.largeSUSYlimit}).}
\label{nondec}
\end{center}
\end{figure}

In general, a non-decoupling behaviour originates 
from a compensation of the mass suppression induced 
through the heavy-particle propagators, by  mass terms
in the interaction vertices. Non-decoupling behaviour has previously been 
analyzed in detail in the case of flavour-preserving MSSM Higgs boson 
decays~\cite{ourHtb,othersHtb,dobado,siannah}.
The non-decoupling contributions to effective FC Higgs Yukawa couplings have 
also been studied in the effective-Lagrangian approach for the quark 
sector~\cite{Demir,dedes} and the lepton sector~\cite{Brignole}.

A condition for non-decoupling of SUSY particles is a common scale
for all SUSY mass parameters. For the simplest assumption, 
the values of the SUSY masses, where all the 
soft breaking squark mass parameters, collectively denoted by $M_{0}$, the 
$\mu$ parameter, the trilinear parameters, collectively denoted by $A$, and 
the gluino mass $M_{\tilde g}$, are chosen to be of the same size and 
much larger than the electroweak scale $M_{EW}$,
\begin{equation}
M_{S} \equiv  M_{0} = M_{\tilde g} = \mu = A  \gg M_{EW}\,.
\label{eq.largeSUSYlimit}
\end{equation}

The numerical results of the analysis are shown in
Fig.~\ref{nondec} for both $h^0,H^0$ production (for $\lambda=0.6$).
One can see that the cross sections tend to a  
non-vanishing value for large values of
$M_S$, a behaviour characteristic of non-decoupling. 
While $M_S$ is displayed only up to $2$ or $3$ TeV, we
have checked that the cross sections remain stable even for very 
large values of $M_S$.

\section{Conclusions}
\label{conclu}

We have considered the production of neutral MSSM Higgs bosons in
association with $b \bar s$ and $s \bar b$ quark pairs. 
We analyzed the flavour
changing effects emanating  from squark-gluino one-loop
contributions, which are the dominant ones in a scenario with
non-minimal flavor mixing in the squark sector.
Both resonant contributions with an intermediate on-shell
$A^0$ boson and non-resonant contributions have been 
investigated.
The highest cross sections for
$e^{+} e^{-} \rightarrow h^0 b \bar s + h^0 s \bar b$ are of
${\cal{O}} (10^{-7})$ pb and of 
${\cal{O}} (10^{-4})$ pb for
$e^{+} e^{-} \rightarrow H^0 b \bar s + H^0 s \bar b$
(and similar for $A^0$) 
which will be too low to
be seen at a future linear collider.

One feature of the SUSY-QCD contributions of general basic interest 
is their non-decoupling behaviour for large values of the SUSY particle
masses. The flavour changing cross sections, when
all soft breaking mass parameters, $\mu$ and
the trilinear parameters, are of the same order of
magnitude and much larger than the electroweak scale,
do not vanish but tend to a constant value.

\section*{Acknowledgments}

This work was supported in part by the European Community's Human
Potential Programme under contract HPRN-CT-2000-00149 
``Physics at Colliders''. 
We are grateful to T. Hahn, S. Dittmaier and S. Heinemeyer
for computing assistance and useful discussions.
We thank P. Slavich for many fruitful discussions and 
for reading the manuscript.
Part of the work of S.P. has been supported by the European Union
under contract No.~MEIF-CT-2003-500030.


\end{document}